\documentclass[
onecolumn,
superscriptaddress,
preprint,
 amsmath,amssymb,
 aps, physrev,
]{revtex4-1}

\usepackage{graphicx}
\usepackage{dcolumn}
\usepackage{bm}
\usepackage{xfrac}
\usepackage{comment}
\usepackage{mathtools}
\usepackage{physics}
\usepackage{makecell}
\usepackage{mathtools}
\usepackage{comment}
\usepackage{physics}
\newcommand*{\kb}{k_{\rm{B}}}

\begin{document}

\preprint{}

\title{\textbf{Quantum dot thermal machines -- a guide to engineering}
}

\author{Eugenia Pyurbeeva}
\email{eugenia.pyurbeeva@mail.huji.ac.il}
\affiliation{The Institute of Chemistry and the Fritz Haber Center for Theoretical Chemistry, The Hebrew University of Jerusalem, Jerusalem 9190401, Israel}
\author{Ronnie Kosloff}
\affiliation{The Institute of Chemistry and the Fritz Haber Center for Theoretical Chemistry, The Hebrew University of Jerusalem, Jerusalem 9190401, Israel}

\date{\today}
\begin{abstract}
Continuous particle exchange thermal machines require no time-dependent driving, can be realised in solid-state electronic devices, and miniaturised to nanometre scale. Quantum dots, providing a narrow energy filter and allowing to manipulate particle flow between the hot and cold reservoirs are at the heart of such devices. It has been theoretically shown that through mitigating passive heat flow, Carnot efficiency can be approached arbitrarily closely in a quantum dot heat engine, and experimentally, values of 0.7$\eta_C$ have been reached. However, for practical applications, other parameters of a thermal machine, such as maximum power, efficiency at maximum power, and noise — stability of the power output or heat extraction — take precedence over maximising efficiency. We explore the effect of internal microscopic dynamics of a quantum dot on these quantities and demonstrate that its performance as a thermal machine depends on few parameters -- the overall conductance and three inherent asymmetries of the dynamics: entropy difference between the charge states, tunnel coupling asymmetry, and the degree of detailed balance breaking. These parameters  act as a guide to engineering the quantum states of the quantum dot, allowing to optimise its performance beyond that of the simplest case of a two-fold spin-degenerate transmission level. 
\end{abstract}
\maketitle

{\let\clearpage\relax
\maketitle}
\section{Introduction}
Quantum thermal machines, heat engines and refrigerators, have been the focus of intense research in the past decades, both as a means of gleaning insight into thermodynamics in the quantum regime \cite{Campbell2025}, and for practical applications. Miniaturisation is a central theme in contemporary technological development, and compact and efficient sources of power and cooling have become an essential requirement for the development of quantum technologies.    

From a practical viewpoint, out of the wide variety of theoretical configurations of quantum thermal machines \cite{Cangemi2024}, and systems in which they have been realised \cite{Myers2022,kosloff2024quantum}: from lasers and solar cells, to superconducting circuits \cite{aamir2025thermally}, qubits, levitated nanoparticles \cite{kuhn2017optically}, and a single atom \cite{Ronagel2016}; for near-term applications the most promising is a continuous electronic particle-exchange heat engine \cite{humphrey2002reversible, Humphrey2005, Kosloff2014}. These devices are autonomous, require no feedback, no time-dependent driving, can be fabricated harnessing existing technological achievements of nanoscale electronics, and readily incorporated with other functional elements.   

The simplest example of such a heat engine is a single electron transistor based on a single quantum dot. Despite its simplicity, such a device has been shown to demonstrate efficiency of over $\eta=0.7$ of the Carnot efficiency \cite{Josefsson2018}, as well as agreement with the Curzon-Ahlborn efficiency \cite{novikov1958efficiency,Curzon1975b} at maximum power. These experimental results follow the theoretical prediction of a system with an infinitely narrow transmission energy bands window optimising thermoelectric efficiency \cite{Mahan1996, Whitney2014}. Generally, SET-based quantum dot heat engines have been predicted to reach thermodynamic limits \cite{Esposito2009, Josefsson2019}.  
 
More recently, a similar experiment with an SET-based heat engine has been performed, employing a single molecule as a working medium \cite{Volosheniuk2025}. It showed comparably high performance, but also a dependence of heat engine operation on the magnetic field -- the suppression of the Kondo effect and the magnetic shift of the energy levels in the molecule reduced both the efficiency of the heat engine and the maximum output power.  

These results demonstrate that for a quantum dot with a more complex internal structure than a single energy level with a two-fold spin degeneracy, this internal structure, or microscopic dynamics: additional spatial degeneracy from symmetry, energy level structure, spin interactions, etc, directly affect its performance as a thermal machine.

This raises a general question which, to our knowledge, has not been considered before -- \emph{how do the internal dynamics of the quantum dot in an SET thermal machine affect its performance, and can it be improved by optimising them?} Or, \emph{what properties should a quantum dot have to make the best-performing thermal machine?} 

For practical operation of a thermal machine, high efficiency is not usually the highest priority, as it typically comes with diminishing power output. Other operation parameters, such as maximum power (maximum cooling power for a refrigerator), efficiency at maximum power, and noise, take precedence. The interplay between the three main parameters of a thermal machine: power, efficiency and constancy (the fluctuations of power output) has been extensively studied \cite{Shiraishi2016, Pietzonka2018}. A \emph{universal trade-off relation} between the three has been proposed \cite{Pietzonka2018}, as well as various schemes attempting to circumvent it and achieve Carnot efficiency at finite power, such as time-dependent cycling \cite{Holubec2018}, and time-reversal symmetry breaking \cite{Benenti2011}, which also suggests the possibility of improvement through modifying internal dynamics of the quantum dot. However, much of the existing theoretical work has been done either completely generally \cite{Shiraishi2016, Pietzonka2018}, or for a special case of a quantum dot with a non-degenerate or a two-fold degenerate transition energy level \cite{Esposito2009, Josefsson2019}. 

We present a systematic study of the effect of internal microscopic dynamics of a quantum dot on its operating parameters as a heat engine for arbitrary dynamics. We note that finding the performance of a known system is always possible. Our aim is the reverse -- to find minimal general guidelines for comparing performances of quantum dots with the internal dynamics unknown. 
  
To limit our considerations to both promising and the experimentally tractable, we consider general devices in the sequential tunnelling regime with the energy level spacing much smaller than $\kb T$, so that the energy transfer window can be considered narrow (the narrow band approximation), in line with the prediction of optimal thermoelectric performance \cite{Mahan1996, Whitney2014} (this regime frequently applies to single-molecule devices \cite{Pyurbeeva2021b}). We also will not consider passive heat flow or vibrational effects. The reason for the latter is that, except for highly specialised cases such as phonon-assisted tunnelling \cite{Sowa2017}, both reduce performance, and are tunable independently of electronic parameters, and thus, for practical applications, can be mitigated separately. 

In the narrow energy band approximation, we treat the unknown internal dynamics of a quantum dot through three asymmetry parameters -- the asymmetry between adding an electron to the QD and removing it from it, relating to the relative number of available quantum states, or the difference in entropy between the charge states, $\Delta S$; the asymmetry between the tunnel couplings of the QD to the baths, $\gamma$; and $\alpha$, charactering the degree of detailed balance breaking in the system. The fourth and final parameter is the overall coupling strength.

To summarise, we demonstrate that in the narrow-band approximation, the dynamics of the system can be described through four characteristic parameters, three asymmetries and an overall scaling. We analyse the performance of an SET based on a quantum dot with non-trivial internal dynamics, including in the case of detailed balance breaking, and demonstrate that power and efficiency properties can be characterised through two parameters -- the coupling strength and a combination of three asymmetries, while noise, or the constancy of power delivery or extraction, depends on all four parameters independency. We believe our results can act as guidelines to engineering quantum states for optimisation of nanoscale thermal machines. 

This paper is structured as follows: in Section \ref{sec-configs}, we describe all configurations of a SET and identify the regimes where it operates as a heat engine and refrigerator; in Section 3, we derive the expressions for efficiency and power of an SET heat engine and refrigerator in the linear regime; in Section 4, we consider, in order, the effect of the entropy difference between the two charge states involved in transport on thermal machine operation, the same effect in the case of normalised conductance, and the effect of detailed balance breaking. Finally, before summarising in Discussion, in Section 5, we study the effects of the entropy difference, tunnel coupling asymmetry, and detailed balance breaking on the noise in the system.

\section{Thermal machine configurations}
\label{sec-configs}
\begin{figure}[h]
\includegraphics[width=\linewidth]{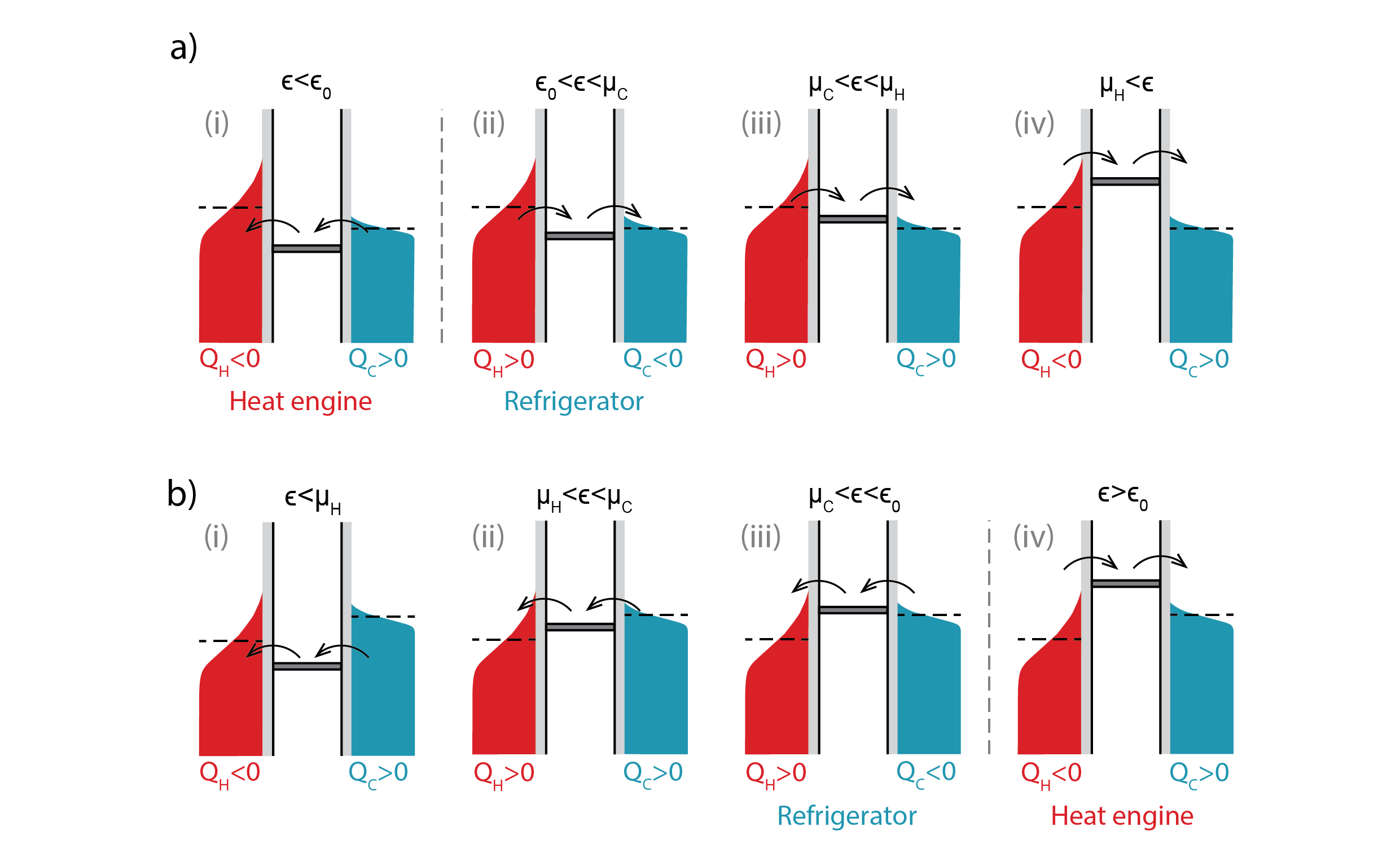}
\caption{All possible configurations of a SET and corresponding thermal operation regimes. a) $\mu_H>\mu_C$, b) $\mu_C>\mu_H$. Current direction (dominant tunnelling direction) is shown with black arrows, and the signs of heat changes of the baths marked beneath them. (Positive sign means heat being deposited in the bath).   \label{fig-configs}}
\end{figure} 

The thermal device we study consists of a single-electron transistor with a single Coulomb-blocked quantum dot with an addition energy $\epsilon=E_{N+1}-E_{N}$ (where $E_{N+1/N}$ is the energy of the $N+1/N$ charge state, the two charge states involved in transport) coupled through tunnel junctions to two  
fermionic heat reservoirs characterized by temperature $T_{H/C}$ and chemical potential
$\mu_{H/C}$.
Figure \ref{fig-configs} shows all possible configurations of a single-electron transistor in respect to its thermal operation regimes. In panel (a), the chemical potential of the hot bath is higher than the chemical potential of the cold bath: $\mu_H>\mu_C$; and in panel (b), the reverse: $\mu_C>\mu_H$. Current direction and the signs of heat arriving into the baths are shown in the figure.  

The direction of heat flow into each bath changes each time the energy level passes through its chemical potential. Electrons arriving into a bath above the chemical potential increase the average energy and correspond to heat flow into the bath (positive in Fig.\ref{fig-configs}), while those arriving beneath the chemical potential lower the mean energy and lead to heat flow form the bath, or negative heat in Fig.\ref{fig-configs}. The opposite holds for electrons leaving the bath.   

It can be noted that there is a direct correspondence between regimes (i)-(iv) in Fig.\ref{fig-configs}a and the regimes (iv)-(i) (in the reverse order) in Fig.\ref{fig-configs}b -- each regime of thermal operation with $\mu_H>\mu_C$ has a corresponding one in $\mu_C>\mu_H$, with the opposite transition energy (low for a heat engine in Fig.\ref{fig-configs}a(i) and high for a heat engine in Fig.\ref{fig-configs}b(iv)). This is due to electron vs. hole current symmetry.

The most significant parameter distinguishing SET the operation regimes in Fig.\ref{fig-configs} is $\epsilon_0$ -- the energy of the transition level corresponding to current reversal, or $I=0$. One way of finding it is to notice that no preferred direction for the current means a zero entropy change when an electron is transferred across the device. As during an electron transfer from one bath to another the entropy change of the quantum dot with electron addition is cancelled out, 

\begin{equation}
\label{eq-dSu}
  \Delta S_u=-\left( \frac{\epsilon_0-\mu_H}{T_H}\right) +\left(\frac{\epsilon_0-\mu_C}{T_C} \right)=0
\end{equation}
where $\Delta S_u$ is the total entropy change of the universe (that of the QD and both baths), $\mu_{H/C}$ and $T_{H/C}$ are the chemical potentials and temperatures of the hot/cold baths, and $\epsilon$ is the charging energy corresponding to current reversal. This gives the current reversal level energy \cite{humphrey2002reversible}:
\begin{equation}
    \epsilon_0=\frac{\mu_C T_H-\mu_H T_C }{T_H-T_C}.
\end{equation}

Note that Eq.\ref{eq-dSu} is identical to the condition of equal populations of the baths at the transition energy level. 

Another point worth noting is that the current reversal point $\epsilon_0$ separates the heat engine and refrigerator operation regimes for both bias voltage directions (Fig.\ref{fig-configs}a(i-ii) and b(iii-iv)). This is in line with the notion that a refrigerator is a heat engine run in reverse -- an infinitesimally small change in $\epsilon$ around $\epsilon_0$ changes the direction of current flow and the device from a heat engine to a refrigerator. 

\section{Thermal machines: efficiency and power}
\subsection{Heat engine}
\label{sec-heat}
\begin{figure}[h]
\includegraphics[width=\linewidth]{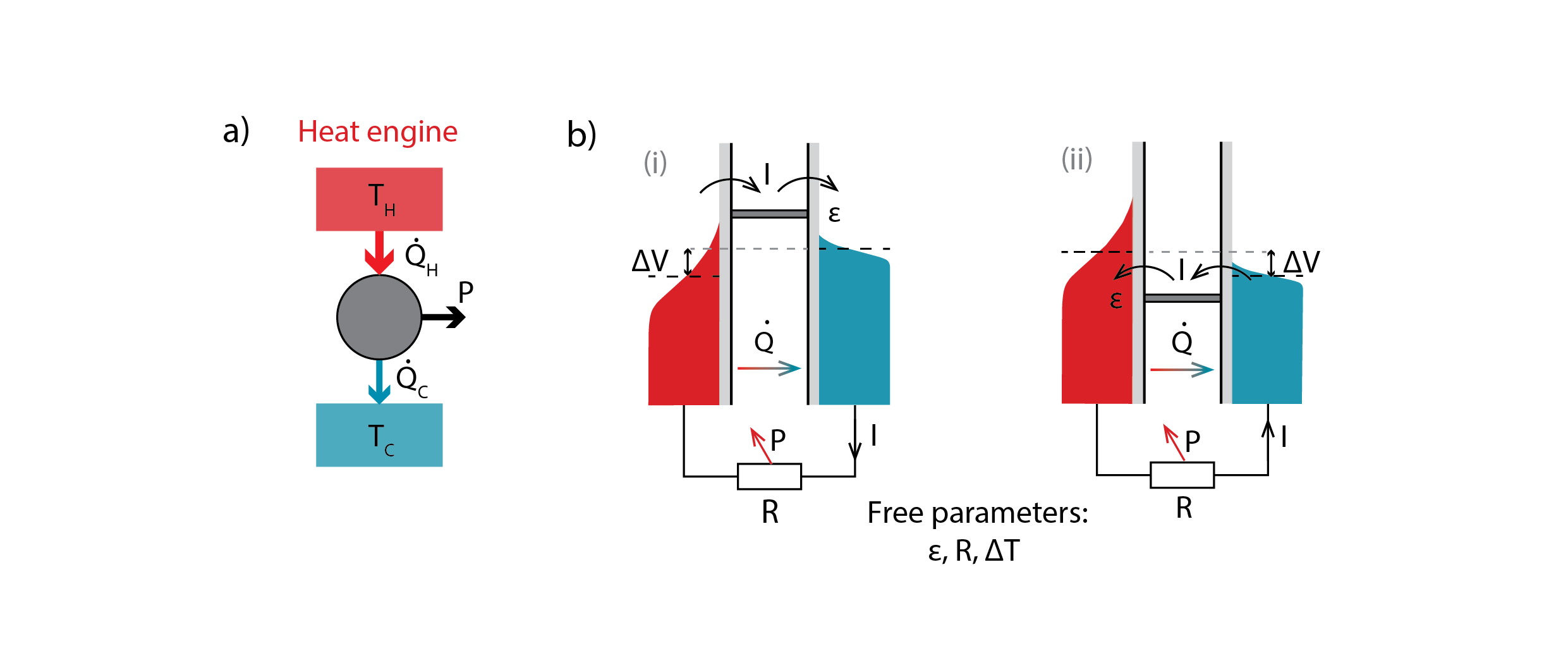}
\caption{a) The general energy flow diagram for a heat engine. b) Two configurations of a SET heat engine, with electron (i) and hole (ii) dominated transport. \label{fig-HE}}
\end{figure} 
The first step in analysing the performance of an SET as a thermal machine is determining the heat and energy exchanges associated with electron transfer events.

Adding an electron with addition energy $\epsilon$ to a thermal bath brings in the amount of heat equal to $\epsilon-\mu$, where $\mu$ is the chemical potential of the bath. Thus, adding an electron above the chemical potential level generates heat in the bath, while in an electron added below the chemical potential leads to the cooling of the bath. The reverse applies for an electron leaving the bath. 

For a heat engine in the electron-dominated SET configuration (Fig.\ref{fig-HE}b(i)) for single electron transfers, $Q_H=\epsilon-\mu_H$, $Q_C=\epsilon-\mu_C$. The heat currents, $\dot{Q_{H/C}}$, indicated in Fig.\ref{fig-HE}a, are therefore, equal to $I Q_{H/C}$ (and are considered positive), where $I$ is the particle current.

For a hole-dominated heat engine (Fig.\ref{fig-HE}b(ii)) the values of heat for a single electron transfer are opposite in sign, but this is compensated by the opposite current direction, leaving the magnitude and direction of heat flows (Fig.\ref{fig-HE}a) the same in both cases (Fig.\ref{fig-HE}b(i) and (ii)). 

The work produced in both heat engine configurations by an electron transfer across the SET is equal to $W=V_b=|\mu_H-\mu_C |$, the bias voltage across the device, and is positive, as electrons move up in potential.

The efficiency of the SET as a heat engine is then equal to:
\begin{equation}
\label{eq-HE-eff0}
    \eta=\frac{W}{Q_H}=\frac{V_b}{| \epsilon-\mu_H |}
\end{equation}
However, this expression can not readily be used. The free parameters in operating an SET as a heat engine are: $\epsilon$, the tuning of the transport energy level; $\Delta T$, the operating temperature difference; and $R$, the load resistance (shown as a resistor in Figs.\ref{fig-HE}, but the load can be any element powered by the heat engine); while the bias voltage $V_b$ is determined by the device to satisfy the Ohm's law in the load: $V_b=IR$.

For simplicity of analysis, we will limit our further considerations to the linear regime, where the transport properties of the device can be described by the Onsager matrix. The current through the device has the form:
\begin{equation}
    I=L \Delta T -G \Delta V
\end{equation}
where $G$ is the conductance, $\partial I/\partial V$, and $L$ -- the thermoelectric susceptibility, $\partial I/\partial T$. Using $\Delta V=IR$, we obtain:
\begin{equation}
    I=\frac{L \Delta T}{1+GR}
\end{equation}
For transport through a single energy level, the Onsager coefficients $G$ and $L$ are related as: 
\begin{equation}
    L=\frac{\varepsilon G}{T}
\end{equation} 
where we define $\varepsilon=\epsilon-\mu=(E_{N+1}-E_{N})-\mu$, the offset between the charging energy of the quantum dot and the chemical potential, as in the linear regime we can consider the chemical potentials of both baths to be nearly equal. 

Returning to the efficiency of the SET as a heat engine (Eq.\ref{eq-HE-eff0}) with the new results and definitions:
\begin{equation}
\label{eq-HE-eff1}
    \eta=\frac{\Delta V}{\varepsilon}=\frac{IR}{\varepsilon}=\frac{\Delta T}{T} \frac{GR}{(1+GR)}
\end{equation}
which gives a natural expression for the ratio between efficiency and the Carnot efficiency:
\begin{equation}
\label{eq-HE-eff}
    \frac{\eta}{\eta_C}=\frac{GR}{(1+GR)}
\end{equation}
Equation \ref{eq-HE-eff} brings forward intuitive understanding of the system: at large $GR$ -- large conductance of the QD and large load resistance, leading to the low current limit, the efficiency approaches that of a Carnot engine, while small $GR$ -- small conductance and low load, with large current, the efficiency of an SET as a heat engine is low. 

Finally, in the analysis of a heat engine, we find the expression for power:
\begin{equation}
    P=I^2R=\left(\frac{\Delta T}{T}\right)^2 \left(\frac{\varepsilon G}{1+GR}\right)^2 R
\end{equation}
which will be used in further numerical analysis.

\subsection{Refrigerator}
\label{sec-ref}
\begin{figure}[h]
\includegraphics[width=\linewidth]{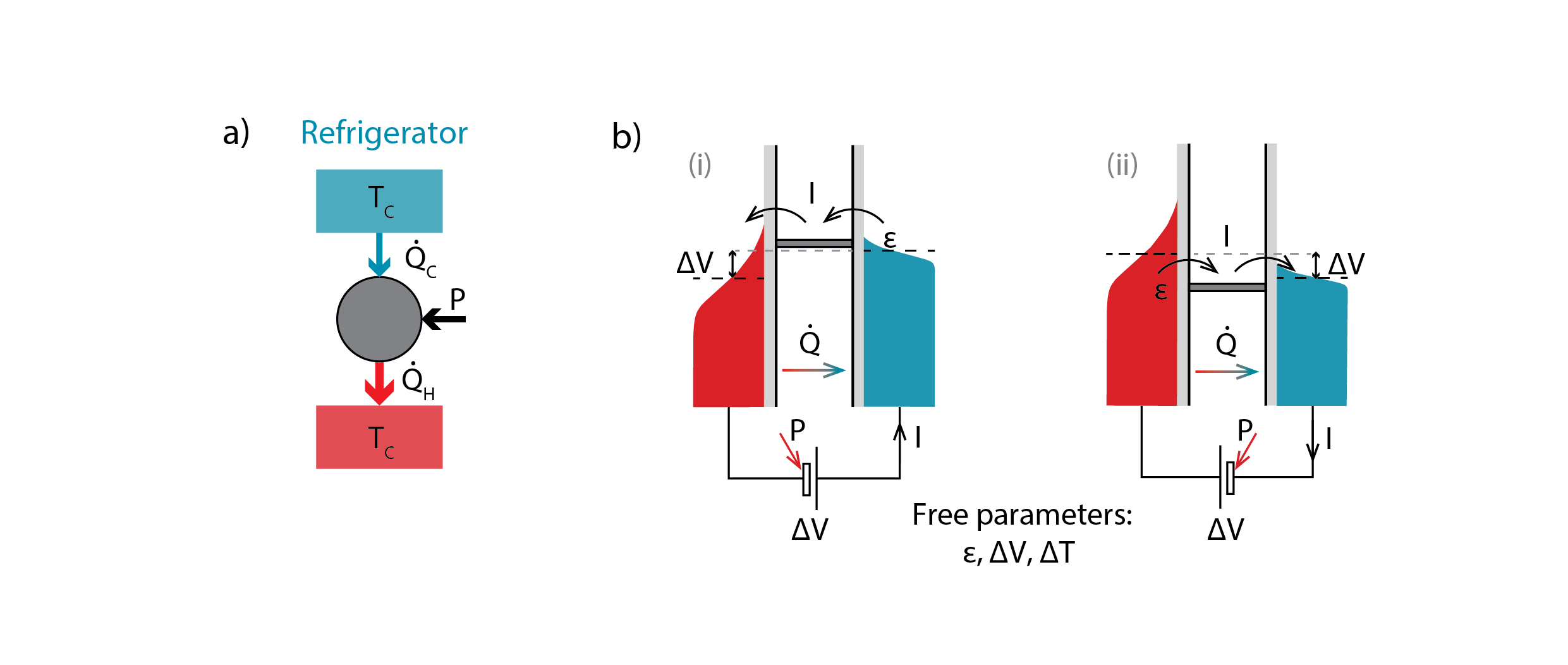}
\caption{a) The general energy flow diagram for a refrigerator. b) Two configurations of a SET operating as a refrigerator, with electron (i) and hole (ii) dominated transport.\label{fig-Ref}}
\end{figure}

It may seem that for an SET to operate as a refrigerator, its parameters need to be precisely tuned. The values of $\epsilon$ required for extracting heat from a cold bath are confined to finite intervals: $\epsilon_0<\epsilon<\mu_C$ or $\mu_C<\epsilon<\epsilon_0$ (Fig.\ref{fig-configs}). From figure \ref{fig-Ref} it can be seen that even illustrating an SET in a refrigerator regime is non-trivial.  The energy window in Fig.\ref{fig-Ref}b(i) in which the population of the cold bath is greater than that of the hot bath and current would flow from the former to the latter is very narrow. This is in contrast to a heat engine, where $\varepsilon$ can be swept from $-\infty$ to $+\infty$, and an SET with a load resistor connected would switch from the hole-dominated to the electron-dominated regime, but remain a heat engine. 
 
However, the tuning requirements for the refrigerator operation regime are less precise than it may seem. Much of it is an illusion due to the fact that for a temperature difference, i.e. difference in the widths of Fermi-distributions, to be visible in an energy diagram, such as Fig.\ref{fig-Ref}, this temperature difference has to be large. 

For an electron-dominated refrigerator (Fig.\ref{fig-Ref}b(i)) the operation conditions are: $\mu_C<\epsilon<\epsilon_0$, and the width of the interval:
\begin{equation}
\label{eq-ref-window}
    \epsilon_0-\mu_C=\frac{\mu_C T_H-\mu_H T_C }{T_H-T_C}-\mu_C=(\mu_C-\mu_H)\frac{T_C}{T_H-T_C}=\frac{\Delta V}{\Delta T}T_C
\end{equation}
In the linear regime, where $T_H\approx T_C \approx T$, and  $\Delta V$ and $\Delta T$ are small, this leads to operation conditions for $\varepsilon=\epsilon-\mu$: $\varepsilon \in \{0; (\Delta V/\Delta T) T \} $, and $\varepsilon$ can be comparable to $T$ is  $\Delta V$ and $\Delta T$ are of similar magnitudes. 

For an SET operating as a refrigerator, the free parameters are $\varepsilon$, the transport energy level, and $\Delta V$ and $\Delta T$, the operating voltage and temperature difference. In the linear regime, the current is:
\begin{equation}
    I=L \Delta T - G \Delta V = \frac{\varepsilon G}{T} \Delta T -G \Delta V
\end{equation}
The heat current $\dot{Q}$ is equal to $\varepsilon I$, which gives the refrigeration efficiency:
\begin{equation}
    \nu=\frac{\dot{Q}}{P}=\frac{\varepsilon I}{I \Delta V}=\frac{\varepsilon}{\Delta V}
\end{equation}
It is noteworthy that this efficiency does not depend on any internal properties of the QD (which are contained in $G(\varepsilon)$), or even $\Delta T$. Also, if $\varepsilon=0$, $\nu=0$, while at $\varepsilon=\varepsilon_0$, $\nu=T/\Delta T$, which agrees with operation at the Carnot regime -- for a Carnot heat engine $\eta_C=W/\dot{Q}_H=\Delta T/T$, while for a refrigerator, $\nu=\dot{Q}_C/P=(\dot{Q}_H-P)/P=1/\eta_C-1=T_C/\Delta T$.

While the refrigerating efficiency depends only on the bias voltage and transport energy level, the cooling power includes dependence on both the QD internal dynamics ($G$) and the temperature difference:
\begin{equation}
    P=\varepsilon I=\frac{\varepsilon^2 G}{T} \Delta T -\varepsilon G \Delta V
\end{equation}
This will be numerically studied below.

\section{The effect of quantum dot dynamics}
\subsection{Entropy difference}
\label{sec-dS}
\begin{figure}[h]
\includegraphics[width=\linewidth]{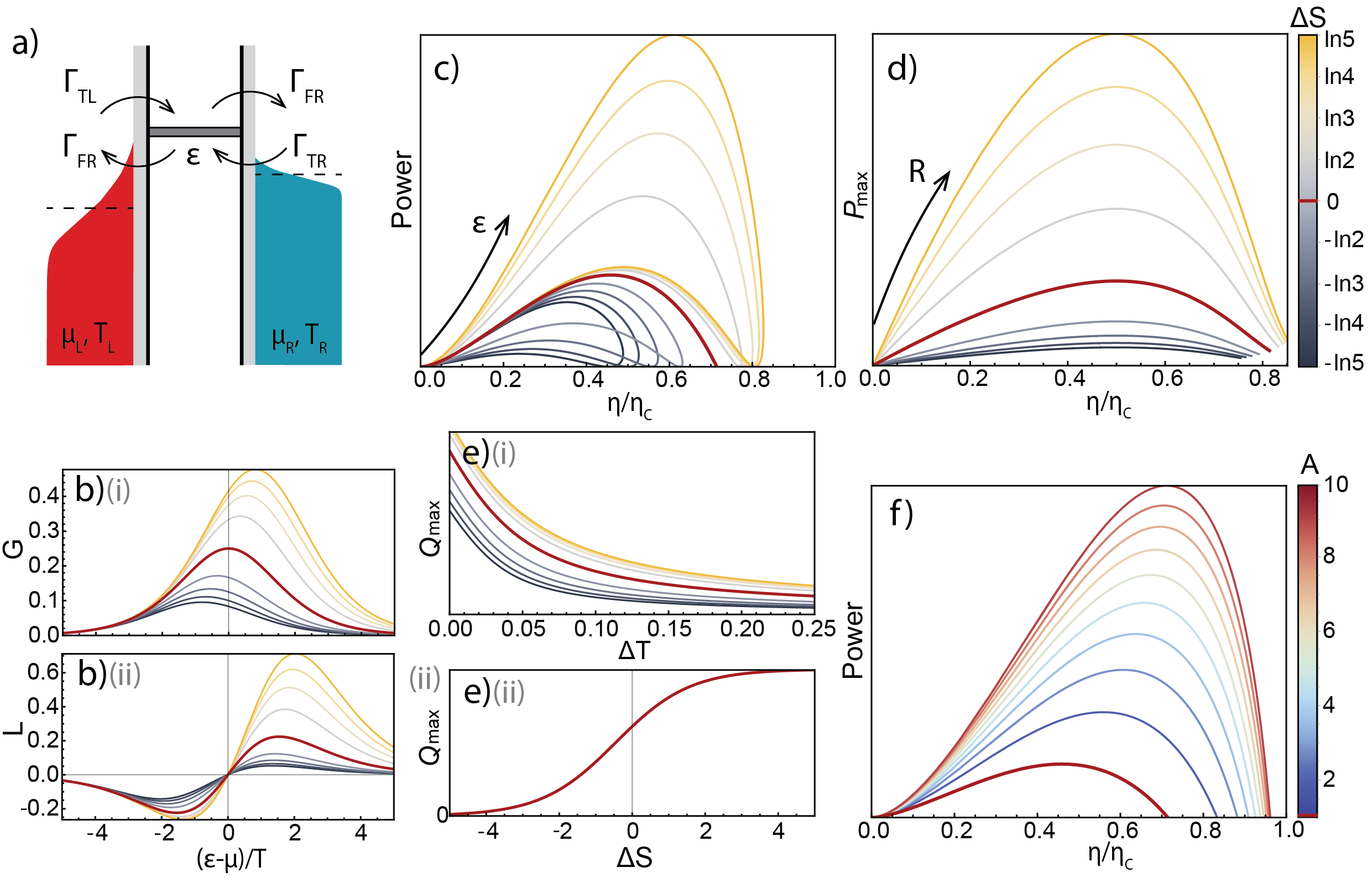}
\caption{a) A general energy diagram for transport through a single energy level, with the rates involved labelled. b) The gate-dependent conductance (i) and thermoelectric susceptibility for several values of $\Delta S$. c) The power-efficiency diagrams for various values of $\Delta S$. The non-degenerate case is shown in red. d) The parameter plots of maximum power vs. efficiency at maximum power at varying load resistances for several values of $\Delta S$. e) (i) The maximum cooling power of an SET refrigerator as a function of temperature difference for several values of $\Delta S$. (ii) Cooling power of an SET refrigerator as a function of $\Delta S$ at fixed $\dd V$ and $\dd T$. f) The power-efficiency diagrams for a non-degenerate QD for increasing values of a coupling strength parameter, $A$ (see Eq.\ref{eq-G-unnorm}).  \emph{As the absolute magnitudes of power and efficiency depend on many parameters: $\Delta T$, $A$, $R$, $\Delta V$, etc, the plots are designed to illustrate the qualitative dependence on the parameters  and thus do not show absolute units.} }.
\label{fig-deg}
\end{figure}

If electron transport is mediated by a single narrow energy band with no additional dissipation, it can be fully described by four exchange rates between the QD and the electrodes (Fig.\ref{fig-deg}a). These rates contain all the information on the internal dynamics of the quantum dot that affects transport, and depend on the densities of states of the electrodes at a single value of $\varepsilon$. We will use them to find $G(\varepsilon)$, the conductance of the QD, and with it, its efficiency and power operating as a heat engine and refrigerator, using the results from the previous section. 

The four rates in question can be written as:
\begin{equation}
\label{eq-rates}
    \begin{dcases}
    \Gamma_{TL}=\gamma_L d_T f_L(\varepsilon)\\
    \Gamma_{FL}=\gamma_L d_F \left(1-f_L(\varepsilon) \right)\\
    \Gamma_{TR}=\gamma_R d_T f_R(\varepsilon)\\
    \Gamma_{FR}=\gamma_R d_F \left(1-f_R(\varepsilon) \right)
    \end{dcases}
\end{equation}
where the directions of associated processes are shown in Fig. \ref{fig-deg}a, the indices $L/R$ (Left/Right) denote the bath involves in the exchange, and $T/F$ (To/From to QD) the direction of the electron exchange. The coefficients $\gamma_{L/R}$  represent the tunnel coupling strengths to the Left/Right bath, $f_{L/R}(\varepsilon)$ are the Fermi-distributions of the bath, where $f(\varepsilon)$ is proportional to the number of electrons with the necessary energy to tunnel into the QD, while $1-f(\varepsilon)$ is the number of available states for an electron to tunnel into the bath. 

The additional coefficients $d_{T/F}$ (corresponding to tunnelling To or From the QD) describe the internal dynamics of the quantum dot. In the case of a simple two-fold spin degeneracy, one of $d_{T/F}$ is equal to 1 and the other to 2, depending on the parity of the occupation of the quantum dot \cite{Harzheim2020}. More generally, from the non-equilibrium fluctuation theorem\cite{Seifert2005, Seifert2012} applied to the SET\cite{Pekola2015}, it can be shown that for two charge states with small energy splitting between the microstates, $\ln(d_T/d_F)=\exp(\Delta S)$, where $\Delta S$ is the entropy differences between the charge states \cite{Pyurbeeva2023}.

The relation between the rate coefficients and entropy difference between the charge states involved in conductance allows to write the Onsager coefficients, conductance and thermoelectric susceptibility, in the form\cite{Pyurbeeva2021b}:
\begin{equation}
\label{eq-GL}
    \begin{dcases}
        G=\frac{1}{T} \frac{\gamma_L \gamma_R}{\gamma_L+\gamma_R}d_T f(\varepsilon) \left( 1-f(\varepsilon-T\Delta S) \right)=\frac{1}{T} \frac{\gamma_L \gamma_R}{\gamma_L+\gamma_R}d_F f(\varepsilon-T\Delta S) \left( 1-f(\varepsilon) \right)\\
        L=\frac{\varepsilon}{T^2} \frac{\gamma_L \gamma_R}{\gamma_L+\gamma_R}d_T f(\varepsilon) \left( 1-f(\varepsilon-T\Delta S) \right)=\frac{\varepsilon}{T^2} \frac{\gamma_L \gamma_R}{\gamma_L+\gamma_R}d_F f(\varepsilon-T\Delta S) \left( 1-f(\varepsilon) \right)
    \end{dcases}
\end{equation}
(See expression derivations in Appendix \ref{app-GL}).

Equations \ref{eq-GL} immediately show that the individual values of $\gamma_L$ and $\gamma_R$ (the tunnel coupling asymmetry) do not play a role, and instead both come in as a single coefficient $\gamma_R\gamma_L/(\gamma_R+\gamma_L)$ -- the harmonic mean. 

Next, we study the effect of $\Delta S$ on the operation of an SET as a heat engine and refrigerator, by analysing the efficiency and power derived in the above sections \ref{sec-heat},\ref{sec-ref} using 
\begin{equation}
\label{eq-G-unnorm}
    G=A f(\varepsilon-T\Delta S) \left( 1-f(\varepsilon) \right)
\end{equation} 
as the characteristic of the device, where $A$ is the ``amplitude'' coefficient, in the case the rates defined in Eq.\ref{eq-rates}, $A=(d_F/T)\cdot\gamma_R\gamma_L/(\gamma_R+\gamma_L) $ (Eq.\ref{eq-GL}). 

The conductance and thermoelectric susceptibility $G$ and $L$ are shown in Fig.\ref{fig-deg}b for a fixed $A$ and varying $\Delta S$. It can be seen that increasing the absolute value of $\Delta S$ leads to a shift of the conductance peak from 0, as well as an asymmetry of the $L$-curve. 

Fig.\ref{fig-deg}c shows the standard power-efficiency plots for an SET heat engine with varying values of $\Delta S$. It should be noted that, unlike the typical diagrams, where the ``loop'' \cite{Josefsson2018} tends to zero after reaching maximum efficiency at non-zero power, here, due to the absence of heat leakage, passive heat flow between the hot and cold baths in the model, maximum efficiency is reached at zero power, and the familiar ``loop'' shape for each non-zero value of $\Delta S$ is made up of the different magnitudes of the positive and negative sections of the thermocurrent $L$. 

It can be seen that increasing $\Delta S$ leads to an increase in both power and efficiency throughout the range of gate tuning $\varepsilon$ (Fig.\ref{fig-deg}c), leading to greater maximum power and maximum efficiency. Figure \ref{fig-deg}d shows maximum power plotted against efficiency at maximum power for a range of load resistances, and demonstrates that $\Delta S$ increases maximum output power at all heat engine operation regimes. 

Finally, while the efficiency of an SET as a refrigerator does not depend on the quantum dot dynamics, the cooling power does, and Fig.\ref{fig-deg}e(i) shows the dependence of maximum cooling power over the range of $\varepsilon$ tunings for a given bias voltage as a function of temperature -- it can be seen that higher entropy difference values consistently lead to higher achievable cooling powers. Fig.\ref{fig-deg}e(ii) demonstrates the same effect by showing the dependence of maximum cooling power of a QD refrigerator with set $\Delta V$ and $\Delta T$ as a function of $\Delta S$, showing that cooling power is also monotonously increased with $\Delta S$. 

The above suggests that a large entropy difference between the two charge states involved in conductance make for overall better operation of an SET as a thermal machine, however, there are further considerations to be taken into account. In the standard form of the rates (Eq.\ref{eq-rates}), there is no formalised distinction between the tunnel rates ($\gamma$'s) and the dynamic coefficients $d$'s. At the same time, one of $d_T$ and $d_F$ is included in the prefactor $A$ in the characteristic conductance (Eq.\ref{eq-G-unnorm}). This is to be expected, as a change of entropy difference between the charge states leads to a change of the entropy of one or both, in turn leading to the change of available states for electron transport, and therefore a change of conductance. It can be seen in Fig.\ref{fig-deg}b(i) -- not just the position of the conductance peak, but its overall magnitude changes with $\Delta S$. 

In reality, it is nearly always impossible to change the entropy difference $\Delta S$ without changing the tunnel couplings and, more generally, $A$. The effect of increasing $A$ for an SET heat engine with a non-degenerate energy level is shown in Fig.\ref{fig-deg}f -- it also leads to higher output power and higher efficiency.   

In the following section we attempt to disentangle the effects of purely geometric coupling strengths and degeneracies (entropy difference).

\subsection{Normalised conductance}
\label{sec-norm}
\begin{figure}[h]
\includegraphics[width=\linewidth]{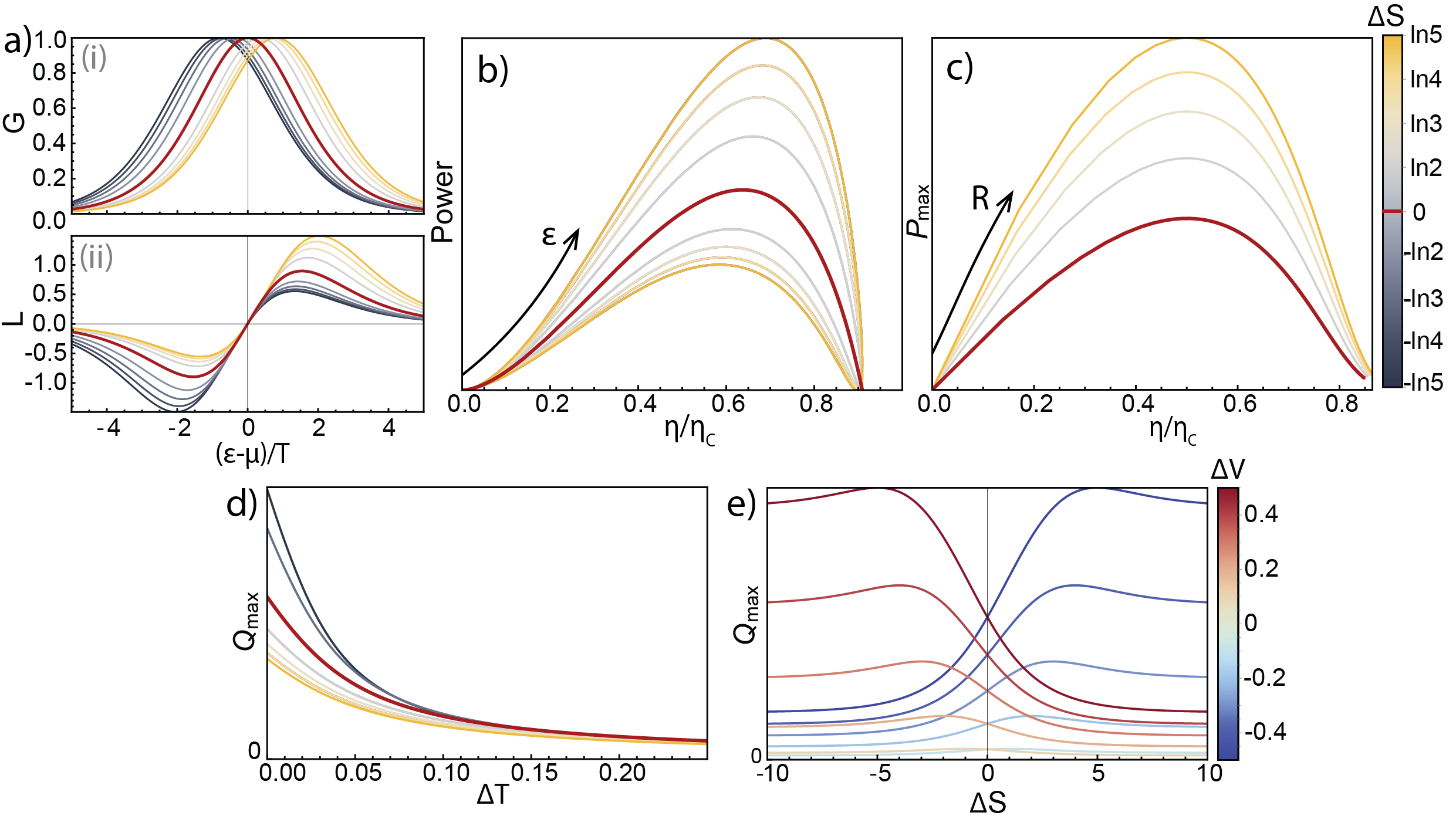}
\caption{a) Normalised gate-dependent conductance (i) and thermoelectric susceptibility for several values of $\Delta S$. b) The power-efficiency diagrams for various values of $\Delta S$ (the positive and negative values of $\Delta S$ have the same dependence). The non-degenerate case is shown in red. c) The parameter plots of maximum power vs. efficiency at maximum power at varying load resistances for several values of $\Delta S$. d) The maximum cooling power of an SET refrigerator with normalised conductance as a function of temperature difference for several values of $\Delta S$. e) Cooling power of an SET refrigerator with normalised conductance as a function of $\Delta S$ at fixed $\Delta T$ and $\Delta V$. \emph{Similarly to Fig.\ref{fig-deg}, the plots are designed to illustrate the qualitative behaviour and  thus the
their scale can very. }
\label{fig-norm}}
\end{figure}
From Fig.\ref{fig-deg}b(i) it can be seen that the entropy difference between the charge states does not significantly affect the shape of the conductance peak, mainly changing its magnitude and position. This is in line with the intuitive notion that for transport through a single narrow energy level, conductance should present a peak of a characteristic width of $\kb T$, located around $\varepsilon=0$, which is what we observe. 

The location of the conductance peak, which in Eq.\ref{eq-GL} is at $\varepsilon=T \Delta S/ 2$ determines the shape of the thermoelectric susceptibility dependence, as $L(\varepsilon)=\varepsilon G(\varepsilon)/T$. As such, the expression:
\begin{equation}
\label{eq-Gnorm}
    G=A f(\varepsilon-\sigma) (1-f(\varepsilon))
\end{equation}
where $A$ and $\sigma$ are effectively fitting parameters quantifying the height of the conductance peak and the thermocurrent asymmetry. This form of $G(\varepsilon)$ and the respective $L(\varepsilon)$ can provide a general description of transport properties of an arbitrary nanodevice with a narrow transmission band -- such an approach had been used in \cite{Volosheniuk2025} to describe an SET heat engine where both conductance and thermocurrent were significantly affected by the Kondo effect and therefore could not be readily described by the simple rate equation approach. The asymmetry parameter $\sigma$ no longer has an immediate physical interpretation in this case.

Here, in order to separate the effects of coupling strengths and entropy difference, we look at the operation of an SET as a thermal machine as a function of $\Delta S$ when the conductance peaks are normalised to the same peak value -- see Fig.\ref{fig-norm}a(i). The $L$-curves in this case still show increased asymmetry in $\Delta S$, however, compared to the non-degenerate curve (shown in red in Fig.\ref{fig-norm}a(ii)) one sign of $\varepsilon$ shows increased $L$, while the other -- a decrease in value. 

For normalised conductances, the power-efficiency plots do not depend on the sign of $\Delta S$, as positive values of $\varepsilon$ for an electron-dominated configuration are exactly symmetrical to negative values of $\varepsilon$ for a hole-dominated device. Fig.\ref{fig-norm}b shows increased power compared to a non-degenerate transmission energy level with an increase of absolute value of $\Delta S$, as well as a negligible reduction of the maximum efficiency around zero power. The increased power follows from the increased values of $L$ at one sign of $\varepsilon$. Fig.\ref{fig-norm} shows that even in the absence of the degeneracy advantage to conductance, the devices with a greater absolute value of $\Delta S$ between the charge states show increased maximum power for the entire range of load resistances. 

For an SET operating as a refrigerator, however, the dependence on $\Delta S$ is more complex. The symmetry between positive and negative values of $\Delta S$ is no longer present, as, unlike a heat engine, where the bias voltage is determined by the tuning of the energy level, and changes sign as current goes through zero, for a refrigerator, the direction of the bias voltage is set externally, thus creating a distinction between electron- and hole-dominated configurations. 

Figure \ref{fig-norm}d shows maximum cooling power over the refrigerating range of $\varepsilon$ values for $\Delta V=0.1\kb T$ and several values of $\Delta S$. Compared to the non-normalised case (Fig.\ref{fig-deg}e(i)), the advantage gained in comparison to $\Delta S=0$, especially at low $\Delta T$ is greater (and it is natural that for a hole-dominated configuration as in Fig.\ref{fig-configs}a(i)), the greatest advantage is at negative $\Delta S$, however, it can be seen that the effect of the entropy difference is less monotonous than in the non-normalised case (Fig.\ref{fig-deg}e(i)). 

To explore this effect, in Fig.\ref{fig-norm}e, we plot the dependence of maximum cooling power at a function of $\Delta S$ for different values of operating bias voltage. The entropy sign opposite to the applied bias dominates, meaning that for hole-dominated configuration (Fig.\ref{fig-Ref}b(ii)), negative values of $\Delta S$ are preferable, while for an electron-dominated configuration (Fig.\ref{fig-Ref}b(i)) the opposite holds true. However, the functions $Q_{max}(\Delta S)$ show distinct maxima, meaning that, unlike the heat engine case, where an increase of $\Delta S$ is always beneficial, even for normalised conductance, for an SET operating as a refrigerator, there is an optimal value of $\Delta S$ that maximises cooling power for given operating conditions. The reason for the existence of an optimum value of $\Delta S$ for a refrigerator lies in the finite range of $\varepsilon$ corresponding to the refrigeration regime of the SET (Eq.\ref{eq-ref-window}). The maximum of the conductance peak, $G(\varepsilon)$ (Eq.\ref{eq-G-unnorm}) lies at $\varepsilon=T\Delta S/2$, and thus, large values of $\Delta S$ can move the conductance peak outside the operation window, leading to a decrease in the device's performance.  
\subsection{Detailed balance breaking}
\label{sec-DB}
Another aspect of the internal dynamics of a quantum dot that can effect its performance as a thermal machine is detailed balance breaking. It has been experimentally observed in a wide variety of physical systems with broken time-reversal symmetry\cite{Entin-Wohlman1995, Nakamura2010, Zhu2014, Khandekar2020, Katcko2019, Chowrira2022}, and has been proved to be compatible with thermodynamics \cite{Alicki2023, Blum2025}.  It has also been shown that in devices based on spintronic spin-selectivity detailed balance breaking leads to an advantage in heat engine performance \cite{Katcko2019, Chowrira2022}. Molecules demonstrating the recently-proposed chirality-induced spin-selectivity effect (CISS) \cite{Naaman2019, Bloom2024} also comprise a promising potential platform for realising novel nanoscale thermal machines breaking detailed balance. 

In order to quantify the effect of detailed balance breaking on thermal machine performance, we introduce a new pair of coefficients $\alpha_+$ and $\alpha_-$ to the rates (Eq.\ref{eq-rates}) to denote a difference between electrons moving in the positive and negative current directions (rightwards vs. leftwards in Fig.\ref{fig-deg}a). The four electron exchange rates are then written as:   

\begin{equation}
\label{eq-rates-DB0}
    \begin{dcases}
    \Gamma_{TL}=\gamma_L \alpha_+ d_T f_L(\varepsilon)\\
    \Gamma_{FL}=\gamma_L \alpha_- d_F \left(1-f_L(\varepsilon) \right)\\
    \Gamma_{TR}=\gamma_R \alpha_- d_T f_R(\varepsilon)\\
    \Gamma_{FR}=\gamma_R \alpha_+ d_F \left(1-f_R(\varepsilon) \right)
    \end{dcases}
\end{equation}

This form of the rates, however, has significant freedom in the distribution of the overall value of the rates between $\gamma$'s, $\alpha$'s, and $d$'s. To remove this uncertainty, we define $d_T=e^{S_1}$, the entropy of the $N+1$ charge state, $d_F=e^{S_0}$, the entropy of the $N$ charge state; $\gamma_L=1+e^\gamma$, $\gamma_R=1+e^{-\gamma}$; $\alpha_+=1+e^{\alpha}$, $\alpha_-=1+e^{-\alpha}$; as well as introduce an overall rate coefficient $\Gamma$, casting the rates in the form:

\begin{equation}
\label{eq-rates-DB}
    \begin{dcases}
    \Gamma_{TL}=\Gamma (1+e^\gamma) (1+e^{\alpha}) e^{S_1} f_L(\varepsilon)\\
    \Gamma_{FL}= \Gamma (1+e^\gamma) (1+e^{-\alpha}) e^{S_0} \left(1-f_L(\varepsilon) \right)\\
    \Gamma_{TR}=\Gamma (1+e^{-\gamma}) (1+e^{-\alpha}) e^{S_1} f_R(\varepsilon)\\
    \Gamma_{FR}=\Gamma (1+e^{-\gamma}) (1+e^{\alpha}) e^{S_0} \left(1-f_R(\varepsilon) \right)
    \end{dcases}
\end{equation}

The expressions for $\gamma_{L/R}$ in Eq.\ref{eq-rates-DB} are inspired by the fact that in the case of a quantum dot with the detailed balance preserved, the conductance and thermoelectric susceptibility (Eq.\ref{eq-GL}) depend on the ``harmonic sum'' combination $\gamma_L \gamma_R/(\gamma_L \gamma_R)$. For $\gamma_L=1+e^\gamma$, $\gamma_R=1+e^{-\gamma}$ this combination is equal to 1, while $\gamma_L/\gamma_R=e^{\gamma}$. The expressions for $\alpha_{+/-}$ were chosen analogously. Thus, the coefficients $\gamma$ and $\alpha$ in Eq.\ref{eq-rates-DB} represent the degree of asymmetry between the left-right tunnel couplings and positive/negative current directions respectively.   

The symmetry breaking between the positive and negative current directions means that charge will be accumulated across the SET, resulting in an induced offset bias voltage in equilibrium (at zero current), $\Delta V$, equal to $-2T\alpha$.

The Onsager coefficients are then found as derivatives from the offset bias, and for the rates given by Eq.\ref{eq-rates-DB}, conductance has the form:
\begin{equation}
\label{eq-G-DB}
    G(\varepsilon)=\frac{\Gamma}{T}e^{S_0} \left(1+\frac{e^{\alpha}+e^{-\gamma}}{1+e^{\alpha-\gamma}} \right) f(\varepsilon-T\Delta S)\left(1-f\left(\varepsilon-T \ln\frac{e^{\alpha}+e^{-\gamma}}{1+e^{\alpha-\gamma}}\right) \right) 
\end{equation}
while, the thermoelectric susceptibility, is, as usual, $L(\varepsilon)=\varepsilon G(\varepsilon)/T$. The derivation for both can be found in Appendix \ref{SI-DB}.

From Eq.\ref{eq-G-DB} it can be seen that the coefficients $\alpha$ and $-\gamma$ act in an identical way. Additionally, if either one of them is equal to zero, then the offset coefficient $\xi=\ln ((e^{\alpha}+e^{-\gamma})/(1+e^{\alpha-\gamma}))$ is also equal to zero, and the expression for conductance is identical to the case of conserved detailed balance (Eq.\ref{eq-GL}). For $\alpha=0$ the statement is trivial, but the case of $\gamma=0$ shows that detailed balance breaking only affects the transport characteristics of an SET if the tunnel couplings are asymmetric. 

The conductance peak occurs at $\varepsilon=(\Delta S+\xi)/2$, meaning that $\sigma=\Delta S+\xi$ acts as an effective entropy, or a generalised asymmetry parameter, to the first order replacing the role of the entropy difference $\Delta S$ in the Onsager coefficients for the case of preserved detailed balance (Eqs.\ref{eq-GL}). The height of the conductance peak depends on $\xi$, but, when normalised, the analysis of the performance of an SET as a thermal machine in the case of detailed balance breaking can follow  Section \ref{sec-norm} with the effective entropy $\sigma$ replacing $\Delta S$. 

This means that tuning the detailed balance breaking and tunnel coupling asymmetry coefficients to increase $\xi$ improves the device's performance as a heat engine. This is in line with the notion that time-reversal symmetry breaking can reduce entropy production \cite{Benenti2017}. The device's performance as a refrigerator can also be optimised, but the optimal value of $\sigma$, as in the previous section, depends of the operating voltage and temperature difference. Finally, despite the arguments for the analysis with normalised conductance, it should also be noted that increasing $\xi$ increases overall conductance and therefore thermal machine performance. 

\section{Noise and constancy}
\begin{figure}[h]
\includegraphics[width=\linewidth]{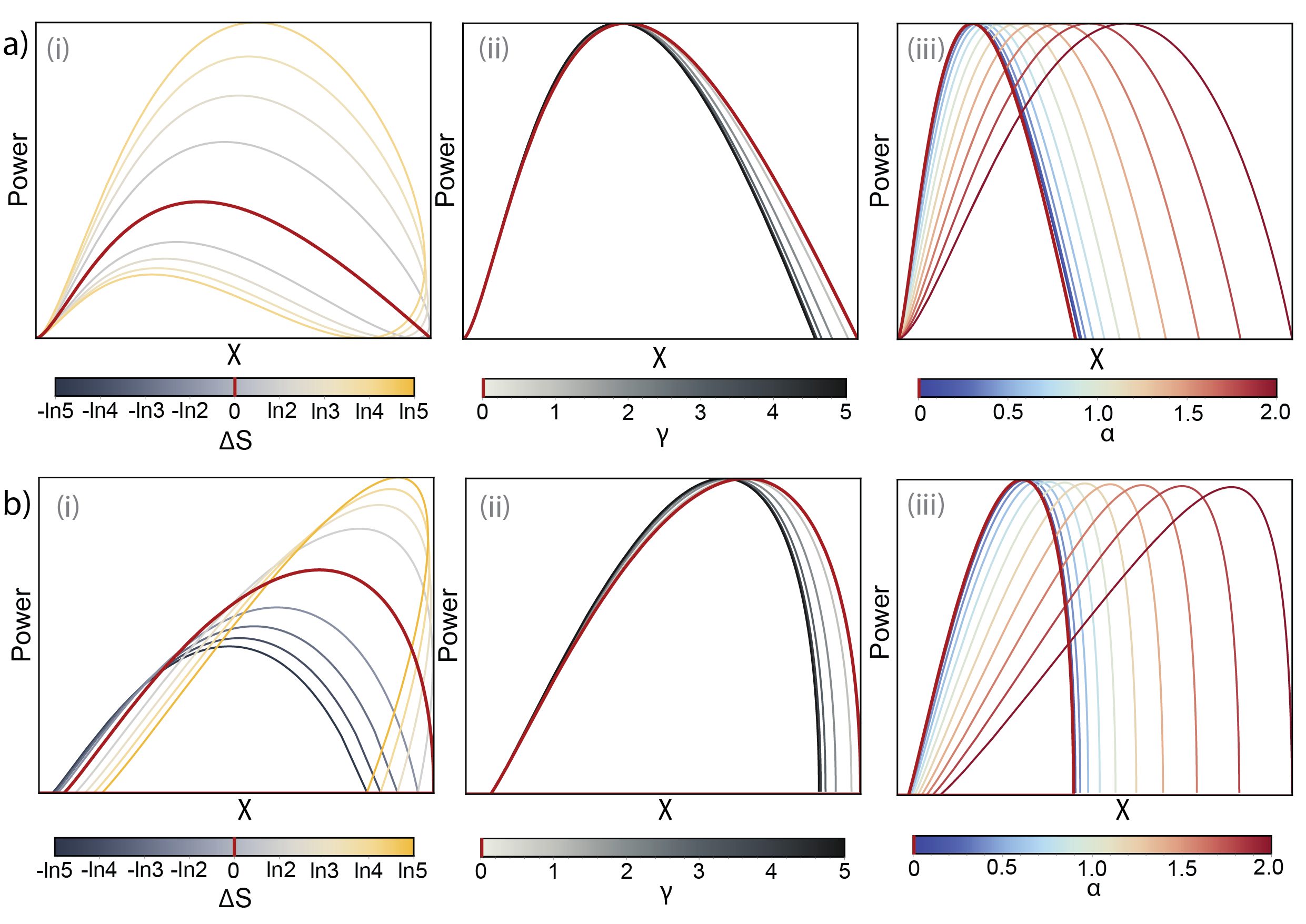}
\caption{a) Power vs. noise (current fluctuations) diagrams for a SET heat engine, shown: (i) for a variety of $\Delta S$ values (the non-degenerate case is shown in red); (ii) for a non-degenerate  transport energy level and  several values of the tunnel coupling asymmetry parameter $\gamma$ (red shows the symmetrically coupled case); (iii) for a non-degenerate transport energy level with $\gamma=0$ and several values of the detailed balance violation parameter $\alpha$ (red shows the case with preserved detailed balance). b) Power vs. noise (current fluctuations) diagrams for a SET refrigerator, shown, similarly (i) for varying values of $\Delta S$; (ii) varying values of $\gamma$; (iii) varying values of $\alpha$.  
\label{fig-noise}}
\end{figure}

The final significant characteristic of a thermal machine that we have not considered so far is noise, or the constancy of power delivery or extraction. The trade-off between power, efficiency, and constancy of a general heat engine has been extensively studied \cite{Shiraishi2016, Pietzonka2018}, however the results are largely based on thermodynamic uncertainty relations (TURs)\cite{Barato2015, Horowitz2020, Seifert2012}, which are called into question in the case of detailed balance breaking \cite{Macieszczak2018, Taddei2023}. Kinetic uncertainty relations (KURs) \cite{Landi2024} based on dynamical analysis hold under detailed balance breaking, providing an independent bound for noise. However, the KUR bound is notoriously weak close to equilibrium \cite{palmqvist2025kinetic}, and therefore not restrictive in the linear response regime.

We thus take a fully dynamical, instead of thermodynamic, approach to the analysis of noise. A general expression for current fluctuations, $\chi$, as a function of four tunnel rates for a SET has the form \cite{Landi2024}:

\begin{equation}
    \chi=\frac{2\Gamma_{TL}\Gamma_{FL}\Gamma_R^2+2\Gamma_{TR}\Gamma_{FR}\Gamma_L^2+4\Gamma_{TL}\Gamma_{FL}\Gamma_{TR}\Gamma_{FR}+(\Gamma_L+\Gamma_R)^2(\Gamma_{TL}\Gamma_{FR}+\Gamma_{TR}\Gamma_{FL})}{\left(\Gamma_L+\Gamma_R \right)^3}
\end{equation}
where, for brevity, we define:  $\Gamma_{L/R}=\Gamma_{TL/R}+\Gamma_{FL/R}$. 

The KUR gives:
\begin{equation}
    \frac{\chi}{I^2}\geq \frac{1}{K}
\end{equation}
where $K$ is activity, equal to $\Gamma_T+\Gamma_F$. Combined with the explicit expression for current through the four rates (Eq.\ref{eq-curr}), this gives a bound:
\begin{equation}
\frac{2\Gamma_{TL}\Gamma_{FL}\Gamma_R^2+2\Gamma_{TR}\Gamma_{FR}\Gamma_L^2+4\Gamma_{TL}\Gamma_{FL}\Gamma_{TR}\Gamma_{FR}+(\Gamma_L+\Gamma_R)^2(\Gamma_{TL}\Gamma_{FR}+\Gamma_{TR}\Gamma_{FL})}{\left(\Gamma_{TL}\Gamma_{FR}-\Gamma_{FL}\Gamma_{TR} \right)^2}\geq 1
\end{equation}

The fluctuations of power output are then equal to $\varepsilon \chi$. However, fluctuations of current in the output circuit are detrimental for thermal machine operation even if they are associated with little power (around $\varepsilon=0$), so for our analysis, we will consider the fluctuations of current directly. 

Unlike the power and efficiency, both based on mean current through the device, in which all three asymmetry parameters: $\Delta S$, $\gamma$, and $\alpha$, came in in a single quantity, noise, or current fluctuations, has a more direct dependence on the microscopic dynamics of the system, and the three asymmetry parameters, $\Delta S$, $\gamma$, and $\alpha$ all affect it differently. Figure \ref{fig-noise}a shows the effect of all three asymmetry parameters on the power vs. noise diagrams for a heat engine for a normalised conductance peak height. 

For a heat engine operating with a non-degenerate transport energy level, at $\varepsilon=\pm \infty$, both power and current fluctuations are equal to zero, as all electron transitions are blocked. At $\varepsilon=0$, in contract, as all transitions are permitted, the noise reaches its maximum, while the output power is zero. In between, the curve reaches a maximum in power, but if low-noise operation is required, this can be achieved with a compromise in power, in line with the general power-efficiency-constancy trade-off rule \cite{Shiraishi2016, Pietzonka2018}.   

A non-zero entropy difference between the charge states, $\Delta S$ splits the dependence for a non-degenerate level into two branches -- one with higher power and one with lower for the same noise levels, corresponding to the asymmetric peaks of the thermocurrent (Fig.\ref{fig-noise}a(i)). The tunnel coupling asymmetry coefficient, $\gamma$, affects the power vs. noise relation is a much less profound way (Fig.\ref{fig-noise}(ii)), as, in the absence of detailed balance breaking, it doesn't affect conductance, and thus output power, but it can be observed that absolute large values of $\gamma$ reduce current fluctuations to a small extent. Finally, the introduction of detailed balance breaking, described by non-zero values of $\alpha$, leads to a dramatic increase of current fluctuations, even in the absence of concurrent tunnel coupling asymmetry and thus the same power output (Fig.\ref{fig-noise}a(iii)). This is in line with the effect of noise amplification with detailed balance breaking, well-known in stochastic thermodynamics \cite{Weiss2003,Weiss2007,Gnesotto2018}.
A non-zero value of $\gamma$ would lead to an advantage in power, similarly to an increase of entropy difference.

For an SET operating as a refrigerator, previously, we had only plotted maximum cooling power over the interval of $\varepsilon$ corresponding to the refrigerator regime, as the cooling efficiency does not depend on $\varepsilon$. However, for a refrigerator, a trade-off exists between cooling power and noise as $\varepsilon$ is varied over its available interval, similarly to a heat engine. Figure \ref{fig-noise} shows the dependence of this trade-off on the asymmetry parameters. For $\Delta S$ below the expected peak values in Fig.\ref{fig-norm}e, increasing the entropy difference leads to a general increase of both power and noise, however, an advantage can be harnessed in extracting greater power at the same noise level. Additionally, as for the ``direct current''-based parameters (Fig.\ref{fig-norm}), unlike the heat engine, the change of sign of $\Delta S$ leads to the opposite effect. The other two asymmetry parameters, $\gamma$ and $\alpha$ affect an SET refrigerator much like a heat engine -- an increase of tunnel coupling asymmetry, i.e. the absolute value of $\gamma$, leads to a slight decrease of current fluctuations, while the introduction of a non-zero $\alpha$, even for a symmetrically coupled QD, leads to a significant increase of noise levels (note that the scale for $\alpha$ in Fig.\ref{fig-noise} is smaller than that for $\gamma$).

\section{Discussion}
\label{sec:discussion}

\begin{figure}
\includegraphics[width=\linewidth]{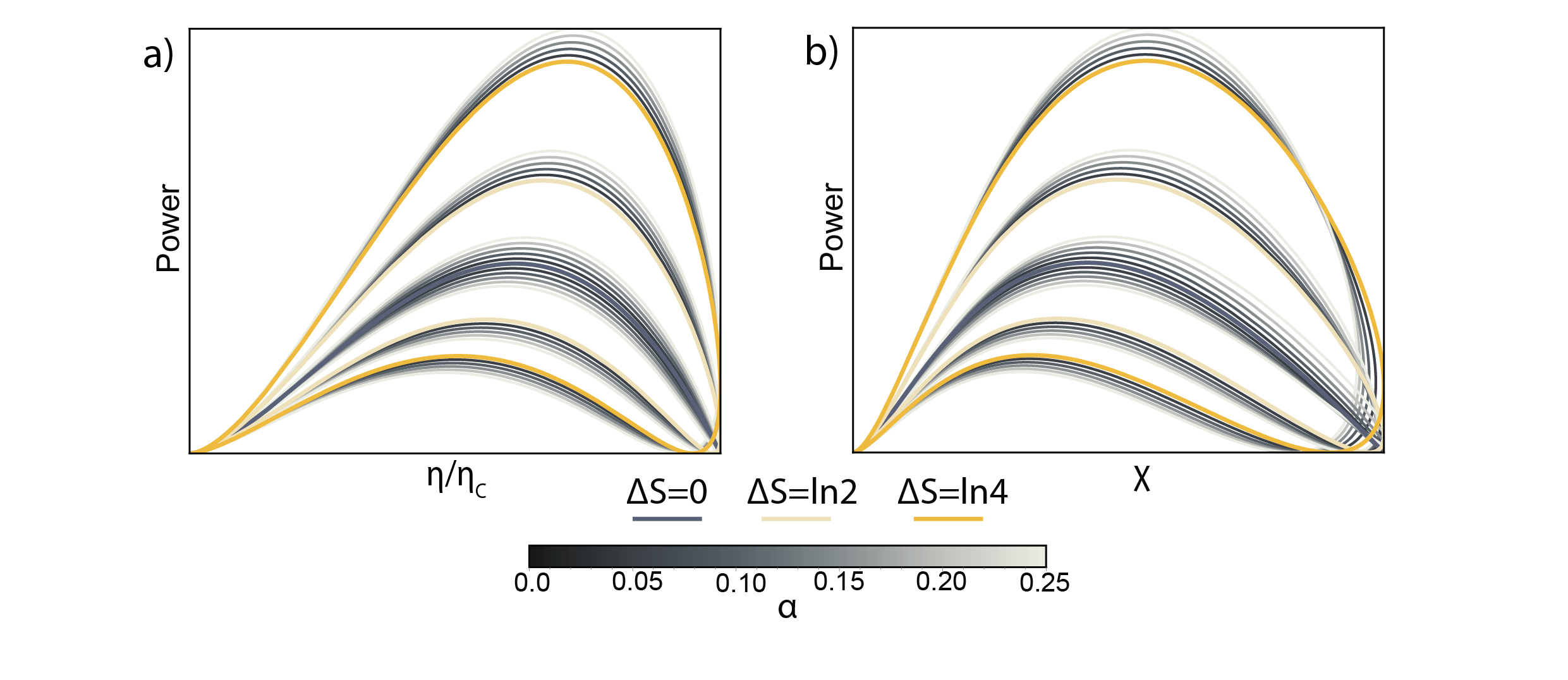}
\caption{a) Power-efficiency diagram for a realistic SET heat engine with large tunnel coupling asymmetry $\gamma=5$ ($\gamma_L/\gamma_R\approx150$) for a non-degenerate transport energy level (grey), two-fold degenerate level (pale yellow) and a four-fold degenerate level (yellow). Grey side lines show the same system with concurrent detailed balance breaking, from $\alpha=0$ to $\alpha=0.25$ (corresponding to the rates ratio of $\approx 30\%$). A power-noise diagram for the same system. 
\label{fig-res}}
\end{figure}

In the work above, we have analysed the dependence of the operation of a single-electron transistor as a thermal machine (heat engine or refrigerator) on a frequently overlooked degree of freedom -- the internal dynamics of the quantum dot. It has been previously demonstrated that changing the dynamics, for instance, by the application of a magnetic field \cite{Volosheniuk2025}, affects the performance of a quantum dot heat engine. This opens a new space for optimisation of nanodevices for future applications, which we explore systematically for the simplest case of a single quantum dot SET. 

In the limit of electron transport governed by a single narrow energy band, which is predicted to optimise performance \cite{Mahan1996, Whitney2014}, the microscopic dynamics of the QD are entirely incorporated into the four electron exchange rates between it and the electrodes (see Fig.\ref{fig-deg}a). Instead of considering the four rates as independent, we perform an effective change of phase-space coordinates and look into their overall magnitude, $\Gamma$, and three asymmetry coefficients: the asymmetry between the probabilities of adding an electron to the QD vs. its removal, described by the entropy difference between the two charge states involved in electron transport, $\Delta S$; the asymmetry between the tunnel coupling strengths of the QD to the two electrodes, $\gamma$; and the asymmetry between positive and negative directions of electron transport, $\alpha$, associated with detailed balance breaking. 

We show that for power and efficiency, the four parameters can be reduced to two: the overall magnitude of the QD conductance, proportional to $\Gamma$, and a combination $\sigma=\Delta S+\ln ((e^{\alpha}+e^{-\gamma})/(1+e^{\alpha-\gamma}))$ (Eq.\ref{eq-G-DB}), which characterises the asymmetry of the thermoelectric susceptibility, $L(\varepsilon)$. Increasing $\Gamma$ necessarily improves the performance of a thermal machine, as it is equivalent to operating multiple machines in parallel. Increasing the thermocurrent asymmetry, $\sigma$, can be achieved either by introducing a difference in entropy between the charge states, or concurrent detailed balance breaking and tunnel coupling asymmetry, and strictly increases the power of an SET operating as a heat engine. For a refrigerator, however, an optimal value of $\sigma$ exists, dependent on its operating parameters: temperature, voltage, and temperature difference. The third significant parameter of a thermal machine, the constancy of its operation, is affected differently by all three asymmetry parameters, and is most significantly increased with detailed balance breaking. 

What do these results mean for optimising a heat engine based on a realistic SET, and how can they be realised? 

Increasing the entropy difference between the charge states, $\Delta S$, has the most profound effect on output power, even when the increase in conductance is corrected for, without a significant change in efficiency or increase in noise. Experimentally, this can be achieved by selecting quantum dots with additional spatial degeneracy -- a dimer molecule has a $\Delta S=\ln4$ for one of the charge transitions \cite{Pyurbeeva2023} (two-fold site degeneracy in addition to spin degeneracy). Other molecules with high degrees of symmetry, and thus spatial degeneracy, such as fullerenes\cite{Kim2014}, can have even higher values of $\Delta S$, which  agrees with their high thermoelectric performance \cite{Kim2014}.     

In experimental settings, SETs typically have high tunnel coupling asymmetry \cite{Pyurbeeva2023}, as small asymmetry in device geometry leads to exponentially great asymmetry in tunnel coupling. This is an advantage, as high values of $\gamma$ decrease noise (see Fig.\ref{fig-noise}a(i)). In solid-state devices
tunnel couplings can be controlled using barrier gates, and a high asymmetry can be set by the experimentalist.

Possible avenues for realising detailed balance violation in a nanodevices include spintronics \cite{Katcko2019, Chowrira2022}, the inclusion of interference, such as Aharonov-Bohm-type rings \cite{Haack2019, Chen2024}, where the asymmetry parameter $\alpha$ could be controlled by applying a magnetic field, or, potentially, CISS \cite{Naaman2019, Bloom2024}. If detailed balance breaking is present, due to the symmetry between $\gamma$ and $\alpha$ in Eq.\ref{eq-G-DB}, a large $\gamma$ will make the effect of even small degrees of detailed balance breaking on the power output of the heat engine significant. 

Figure \ref{fig-res} outlines the realistic advantages that can be garnered in an SET heat engine. It shows the power-efficiency and power-noise diagrams of an SET heat engine with high tunnel coupling asymmetry ($\gamma_L/\gamma_R\approx150$) three different degeneracy values (0, 2, and 4) and an additional small degree of detailed balance breaking (up to $\approx 30\%$). 

We have presented a systematic study of the effect of internal dynamics of a quantum dot on thermal machine performance for the simplest possible configuration, a single quantum dot coupled to two baths. We then identified minimum guidelines for comparing and optimising SET thermal machines with unknown internal dynamics.
This study was carried out under the assumptions: linear response, weak coupling limit, and a narrow transmission energy band,  the simplest regime from the theoretical perspective.

Nevertheless the derived form of $G(\varepsilon)$ and $L(\varepsilon)$ has been experimentally shown \cite{Pyurbeeva2021b, Volosheniuk2025} to hold well in realistic molecular devices with non-trivial structures, even beyond the original domain of the assumptions. For instance, when $\Delta T/T \sim 1$, or when energy the level splitting becomes significant.  This success stems from the generality of the effective fitting of $G(\varepsilon)$ and $L(\varepsilon)$ with Eq.\ref{eq-Gnorm}, which hold even when the fitting parameters lose their physical interpretation\cite{Volosheniuk2025}.

The main aim of the work was to highlight the potential benefit of engineering the internal structure of a quantum dot for optimising device performance. While our results are limited to a simple case there is no reason to believe that the effects of internal microscopic dynamics should be limited to  such cases. Optimising microscopic dynamics has been a long-standing problem in applied thermoelectrics, but has been largely (with rare exceptions \cite{allahverdyan2004maximal,allahverdyan2010optimal}) overlooked in the field of quantum thermodynamics. Related studies include the role of degenerate excited state in an Otto engine \cite{allahverdyan2004maximal,allahverdyan2010optimal}, or, more recently, employing catalytic states to gain advantage in heat engine and refrigerator performance \cite{Biswas2024, Lobejko2025, Fu2025}. 

We believe that introducing quantum dots with non-trivial microscopic dynamics will lead to enhanced performance.  Such complex systems,   with time-dependent driving, multiple quantum dots, non-thermal resources \cite{monsel2025precision}, will be beneficial. These devices go beyond our simple initial assumptions, such as for strong coupling. This topic deserving greater attention, especially as such devices approach practical applicability.  



\vspace{6pt} 








\acknowledgments{We thank Juliette Monsel for useful discussions of noise in quantum dot systems. E.P. is grateful to the Azrieli Foundation for the award of an Azrieli Fellowship.}



\appendix
\section[\appendixname~\thesection]{$G$ and $L$ expression derivation}
\label{app-GL}
Here, we give a slightly modified and abridged derivation of equations \ref{eq-GL}, first derived in  \cite{Pyurbeeva2021b}.

The population of the quantum dot coupled to two baths in the general case is equal to:
\begin{equation}
    P_1=\frac{\Gamma_{TL}+\Gamma_{TR}}{\Gamma_{TL}+\Gamma_{TR}+\Gamma_{FL}+\Gamma_{FR}}
\end{equation}
where we continue with the index notation of Eqs.\ref{eq-GL}. In the equilibrium case, where $f_L(\varepsilon)=f_R(\varepsilon)=f(\varepsilon)$:
\begin{equation}
    P_1=\frac{(\gamma_L+\gamma_R)d_T f(\varepsilon)}{(\gamma_L+\gamma_R)d_T f(\varepsilon)+(\gamma_L+\gamma_R)d_F \left( 1-f(\varepsilon) \right)}=\frac{d_T f(\varepsilon)}{d_T f(\varepsilon)+d_F \left( 1-f(\varepsilon) \right)}
\end{equation}
Expanding the Fermi-distribution into its exponential form, this gives:
\begin{equation}
    P_1=\dfrac{1}{1+\dfrac{d_F}{d_T}e^{\frac{\varepsilon}{T}}}=\dfrac{1}{1+e^{\left(\frac{\varepsilon}{T}-\ln \left(\frac{d_T}{d_F} \right) \right)}}=\dfrac{1}{1+e^{\left(\frac{\varepsilon-T\Delta S}{T} \right)}}=f(\varepsilon-T\Delta S)
\end{equation}
where we use $d_T/d_F=\exp(\Delta S)$ (we set $\kb=1$ throughout). The same result can also be derived from the Maxwell relations, or the Gibbs distribution\cite{Pyurbeeva2020b}, or maximising entropy.

Then,
\begin{equation}
    P_0=1-f(\varepsilon-T\Delta S)
\end{equation}

The above are equilibrium populations. The condition for the QD population to be stationary (even outside of equilibrium) is:
\begin{equation}
\label{eq-st-con}
    \begin{dcases}
        P_0+P_1=1\\
        \frac{\dd P_1}{\dd t}=(\Gamma_{TL}+\Gamma_{TR})P_0-(\Gamma_{FL}+\Gamma_{FR})P_1=0
    \end{dcases}
\end{equation}
Now, we modify the Fermi-distribution of the left bath, so that $f_L(\varepsilon)=f(\varepsilon)+\dd f$, $f_R(\varepsilon)=f(\varepsilon)$, staying in the linear regime, so that $\dd f$ is small. The populations change by $\dd p$: $P_1^{ne}=P_1+\dd p$, $P_0^{ne}=P_0-\dd p$ (satisfying the first condition for non-equilibrium populations); while the hopping rates associated with the exchange with the right bath are not affected. 

The second condition for a stationary state gives, to the linear terms:
\begin{equation}
    \gamma_L d_T \dd f P_0-(\gamma_L+\gamma_R)d_T f(\varepsilon)\dd p +\gamma_L d_F \dd f P_1- (\gamma_L+\gamma_R)d_F\left(1-f(\varepsilon)\right)\dd p=0
\end{equation}
which gives:
\begin{equation}
    \dd p=\frac{\gamma_L}{(\gamma_L+\gamma_R)}\frac{d_T P_0 +d_F P_1}{d_T f(\varepsilon)+d_F \left(1-f(\varepsilon) \right)}\dd f=\frac{\gamma_L}{(\gamma_L+\gamma_R)} \dfrac{P_1+\frac{d_T}{d_F}P_0}{\left(1+\left(\frac{d_T}{d_F}-1 \right)f(\varepsilon)\right)}\dd f
\end{equation}
This, together with certain properties of the Fermi-distribution, mainly $f(x)/(1-f(x))=e^{-x}$, and the explicit form of its derivatives, allows to find the population changes with the application of a potential or temperature change to the left electrode:
\begin{equation}
\label{eq-dp0}
    \begin{dcases}
        \frac{\dd p}{\dd V}=\frac{1}{T}\frac{\gamma_L}{(\gamma_L+\gamma_R)} \dfrac{P_1+\frac{d_T}{d_F}P_0}{\left(1+\left(\frac{d_T}{d_F}-1 \right)f(\varepsilon)\right)} f(\varepsilon) \left(1-f(\varepsilon) \right)
        \\
        \frac{\dd p}{\dd T}=\frac{\varepsilon}{T^2}\frac{\gamma_L}{(\gamma_L+\gamma_R)} \dfrac{P_1+\frac{d_T}{d_F}P_0}{\left(1+\left(\frac{d_T}{d_F}-1 \right)f(\varepsilon)\right)} f(\varepsilon) \left(1-f(\varepsilon) \right)
    \end{dcases}
\end{equation}

This can be further simplified, using
\begin{equation}
        P_1+\frac{d_T}{d_F}P_0=f(\varepsilon-T\Delta S)+e^{\Delta S} e^{\frac{\varepsilon-T\Delta S}{T}}f(\varepsilon-T\Delta S)=f(\varepsilon-T\Delta S)\left(1+e^{\frac{\varepsilon}{T}} \right)=\frac{f(\varepsilon-T\Delta S)}{f(\varepsilon)}
\end{equation}
and 
\begin{equation}
 \frac{1}{1+\left(\frac{d_T}{d_F}-1 \right)f(\varepsilon)}=\frac{1+e^{\frac{\varepsilon}{T}}}{1+e^{\frac{\varepsilon}{T}}+\frac{d_T}{d_F}-1}=\frac{1+e^{\frac{\varepsilon}{T}}}{e^{\frac{\varepsilon}{T}}+e^{\Delta S}}=e^{-\Delta S}\frac{f(\varepsilon-T\Delta S)}{f(\varepsilon)}
\end{equation}
Putting the above together with Eq.\ref{eq-dp0}, gives:
\begin{equation}
    \begin{dcases}
        \frac{\dd p}{\dd V}=\frac{1}{T}\frac{\gamma_L}{(\gamma_L+\gamma_R)}f(\varepsilon-T\Delta S)\left(1-f(\varepsilon-T\Delta S) \right)
        \\
        \frac{\dd p}{\dd T}=\frac{\varepsilon}{T^2}\frac{\gamma_L}{(\gamma_L+\gamma_R)}f(\varepsilon-T\Delta S)\left(1-f(\varepsilon-T\Delta S) \right) 
    \end{dcases}
\end{equation}
Stationary current through the junction right of the quantum dot (in the stationary case currents through both junctions are equal) is equal to:
\begin{equation}
    I=P_1^{ne}\Gamma_{FR}-P_0^{ne}\Gamma_{TR}=(P_1+\dd p)\Gamma_{FR}-(P_0-\dd p)\Gamma_{TR}=(\Gamma_{FR}+\Gamma_{TR})\dd p
\end{equation}
as the equilibrium current is zero. 

The sum of the rates can also be simplified:
\begin{equation}
    \Gamma_{FR}+\Gamma_{TR}=\gamma_R \left(d_T f(\varepsilon)+d_F (1-f(\varepsilon)) \right)=\gamma_R d_T \left(f(\varepsilon)+e^{-\Delta S} e^{\frac{\varepsilon}{T}} f(\varepsilon)\right)=\gamma_R d_T \frac{f(\varepsilon)}{f(\varepsilon-T\Delta S)}
\end{equation}
Similarly, taking $d_F$ outside of the brackets instead
\begin{equation}
    \Gamma_{FR}+\Gamma_{TR}=\gamma_R \left(d_T f(\varepsilon)+d_F (1-f(\varepsilon)) \right)=\gamma_R d_F \left(e^{\Delta S} f(\varepsilon)+(1-f(\varepsilon))\right)=\gamma_R d_F \frac{1-f(\varepsilon)}{1-f(\varepsilon-T\Delta S)}
\end{equation}

The Onsager coefficients we need to obtain can be written as:
\begin{equation}
    \begin{dcases}
        G=\frac{\dd I}{\dd V}=(\Gamma_{FR}+\Gamma_{TR})\frac{\dd p}{\dd V}\\
        L=\frac{\dd I}{\dd T}=(\Gamma_{FR}+\Gamma_{TR})\frac{\dd p}{\dd T}
    \end{dcases}
\end{equation}
Finally, combining the expressions for the change of population with $\Gamma_{FR}+\Gamma_{TR}$, we arrive at the desired results:
\begin{equation}
    \begin{dcases}
        G=\frac{1}{T}\frac{\gamma_L\gamma_R}{(\gamma_L+\gamma_R)}d_T f(\varepsilon)\left( 1-f(\varepsilon-T\Delta S)\right)=\frac{1}{T}\frac{\gamma_L\gamma_R}{(\gamma_L+\gamma_R)}d_F f(\varepsilon-T\Delta S)\left( 1-f(\varepsilon)\right)\\
        L=\frac{\varepsilon}{T^2}\frac{\gamma_L\gamma_R}{(\gamma_L+\gamma_R)}d_T f(\varepsilon)\left( 1-f(\varepsilon-T\Delta S)\right)=\frac{\varepsilon}{T^2}\frac{\gamma_L\gamma_R}{(\gamma_L+\gamma_R)}d_F f(\varepsilon-T\Delta S)\left( 1-f(\varepsilon)\right)
    \end{dcases}
\end{equation}

\section[\appendixname~\thesection]{The modified rates}
The standard form of the electron exchange rates (Eq.\ref{eq-rates}) does explicitly define $\gamma_{T/F}$ and $d_{T/F}$. For a degenerate energy level, it is typical to take $d_{T/F}$ equal to the degeneracies of the charge states\cite{Harzheim2020}, but the only general relation is $d_T/d_F=e^{\Delta S}$, which leaves the overall scaling of the coefficients free. 

A natural way of setting the values of the coefficients is $d_T=e^{S_1}$, $d_F=e^{S_0}$, where $S_{1/0}$ is the entropy of the $N+1/N$ charge state. This agrees with the coefficients being equal to the degeneracy in the case of a degenerate level. This gives the form of the rates:
\begin{equation}
\label{eq-rates-mod}
    \begin{dcases}
    \Gamma_{TL}=\gamma_L e^{S_1} f_L(\varepsilon)\\
    \Gamma_{FL}=\gamma_L e^{S_0} \left(1-f_L(\varepsilon) \right)\\
    \Gamma_{TR}=\gamma_R e^{S_1} f_R(\varepsilon)\\
    \Gamma_{FR}=\gamma_R e^{S_0} \left(1-f_R(\varepsilon) \right)
    \end{dcases}
\end{equation}
With this definition, the conductance and thermoelectric susceptibility in Eq.\ref{eq-GL} can be written in entropic form:
\begin{equation}
    \begin{dcases}
    G=\frac{\bar{\gamma}}{T}\frac{e^{\frac{\varepsilon+T S_0}{T}}}{\left(1+e^\frac{\varepsilon}{T} \right)\left(1+e^\frac{\varepsilon-T\Delta S}{T} \right)}=\frac{\bar{\gamma}}{T}\frac{1}{\left(e^{-\frac{\varepsilon}{2T}}+e^{\frac{\varepsilon}{2T}} \right) \left( e^{-\frac{\varepsilon}{2T}-S_0}+e^{\frac{\varepsilon}{2T}-S_1}\right)} \\
    L=\frac{\bar{\gamma}\varepsilon}{T}\frac{e^{\frac{\varepsilon+T S_0}{T}}}{\left(1+e^\frac{\varepsilon}{T} \right)\left(1+e^\frac{\varepsilon-T\Delta S}{T} \right)}=\frac{\bar{\gamma}}{T}\frac{\varepsilon}{\left(e^{-\frac{\varepsilon}{2T}}+e^{\frac{\varepsilon}{2T}} \right) \left( e^{-\frac{\varepsilon}{2T}-S_0}+e^{\frac{\varepsilon}{2T}-S_1}\right)}
    \end{dcases}
\end{equation}
where $\bar{\gamma}=\gamma_L \gamma_R/(\gamma_L + \gamma_R)$. 

If there is necessity to analyse noise, or constancy, of a thermal machine in addition to the transport and energetic parameters discussed in Section \ref{sec-norm}, it does not suffice to simply normalise conductance, and rates have to be modified directly.  

Conductance, $G(\varepsilon)$ is linear in overall scaling of the rates, therefore in order to normalise all processes by conductance, it is sufficient to divide all rates by the entropy-dependent part of the conductance peak value. Conductance maximum occurs at $\varepsilon=T\Delta S/2$, and is equal to:
\begin{equation}
    G_{max}=\frac{\bar{\gamma}}{T}\frac{e^{S_1}}{\left(1+e^{\frac{\Delta S}{2}} \right)^2}=\frac{\bar{\gamma}}{2T}\frac{e^{\frac{S_1+S_0}{2}}}{\left(1+\sinh(\frac{\Delta S}{2}) \right)^2}
\end{equation}
The fact that the conductance depends not only on the difference between the charge state entropies $\Delta S$, but also on the sum, is natural, as an increased number of total states accessible increases conductance.

For the degeneracy coefficients modified to exclude the dependence of conductance on entropy, this gives us:
\begin{equation}
    \begin{dcases}
    d_T=\left(1+e^{\frac{\Delta S}{2}} \right)^2\\
    d_F=e^{-\Delta S}\left(1+e^{\frac{\Delta S}{2}} \right)^2
    \end{dcases}
\end{equation}
or, in a symmetrised form:
\begin{equation}
    \begin{dcases}
    d_T=2e^{\frac{\Delta S}{2}}\left(1+\sinh(\frac{\Delta S}{2}) \right)^2\\
    d_F=2e^{-\frac{\Delta S}{2}}\left(1+\sinh(\frac{\Delta S}{2}) \right)^2
    \end{dcases}
\end{equation}

\section[\appendixname~\thesection]{Detailed balance breaking, $G$ and $L$}
\label{SI-DB}
The derivation of conductance the thermoelectric susceptibility for the case of detailed balance breaking  largely follows the same logical as that for the standard case (Appendix \ref{app-GL}). 

The population of the quantum dot in the general case is given by:
\begin{equation}
\label{eq-SI-pop}
    P_1=\frac{\Gamma_{TL}+\Gamma_{TR}}{\Gamma_{TL}+\Gamma_{TR}+\Gamma_{FL}+\Gamma_{FR}}
\end{equation}
And the current:
\begin{equation}
\label{eq-curr}
    I=\Gamma_{TL}P_0-\Gamma_{FL}P_1=\Gamma_{FR}P_1-\Gamma_{TR}P_0=\frac{\Gamma_{TL}\Gamma_{FR}-\Gamma_{FL}\Gamma_{TR}}{\Gamma_{TL}+\Gamma_{TR}+\Gamma_{FL}+\Gamma_{FR}}
\end{equation}
We find the equilibrium state of the device by setting $I=0$, or equivalently, $\Gamma_{TL}\Gamma_{FR}-\Gamma_{FL}\Gamma_{TR}=0$. Substituting the rates introduced in Eq.\ref{eq-rates-DB0} into this condition we arrive at:
\begin{equation}
    \alpha_+^2 f_L(\varepsilon)(1-f_R(\varepsilon))=\alpha_-^2 f_R(\varepsilon)(1-f_L(\varepsilon))
\end{equation}
With the properties of the Fermi distribution: $1-f(x)=e^xf(x)$, this reduces to:
\begin{equation}
    \alpha^2_+e^{-\frac{V_R}{T}}=\alpha^2_-e^{-\frac{V_L}{T}}
\end{equation}
or:
\begin{equation}
    \frac{\alpha_+}{\alpha_-}=e^{-\frac{\Delta V}{2T}}
\end{equation}
This shows that the equilibrium state of the device, i.e. the state with zero current, occurs at a finite bias voltage $\Delta V$, equal to:
\begin{equation}
    \frac{\Delta V}{T}=-2\ln \left( \frac{\alpha_+}{\alpha_-} \right)
\end{equation}
The form of $\Delta V$ in a further argument in support of the form of the rate coefficients proposed in Eq.\ref{eq-rates-DB}. There, $\alpha_+/\alpha_-=e^\alpha$, and thus, $\Delta V/T=-2\alpha$.

We divide this induced bias voltage symmetrically, postulating that in equilibrium the Fermi-distributions of the electrodes are: 
\begin{equation}
    \begin{dcases}
    f_L=f(\varepsilon+\alpha T)\\
    f_R=f(\varepsilon-\alpha T)
    \end{dcases}
\end{equation}
Using these equilibrium distributions, and the rate expressions defined in Eq.\ref{eq-rates-DB0} (to keep the expressions concise), we return to Eq.\ref{eq-SI-pop} to find the equilibrium population of the quantum dot:
\begin{multline}
    P_1=\frac{\gamma_L e^{S_1}\alpha_+ f(\varepsilon+\alpha T)+\gamma_R e^{S_1}\alpha_- f(\varepsilon-\alpha T)}{\gamma_L e^{S_1}\alpha_+ f(\varepsilon+\alpha T)+\gamma_L e^{S_0}\alpha_-(1- f(\varepsilon+\alpha T))+\gamma_R e^{S_1}\alpha_- f(\varepsilon-\alpha T)+\gamma_R e^{S_0}\alpha_+ (1- f(\varepsilon-\alpha T))}\\
    =\frac{e^{S_1} \left(\gamma_L \alpha_+ f(\varepsilon+\alpha T)+\gamma_R \alpha_- f(\varepsilon-\alpha T) \right)}{\gamma_L f(\varepsilon+\alpha T) \left(\alpha_+ e^{S_1}+\alpha_+ e^{-\alpha} e^{\frac{\varepsilon+\alpha T}{T}} \right)+\gamma_R f(\varepsilon-\alpha T) \left(\alpha_- e^{S_1}+\alpha_- e^{\alpha} e^{\frac{\varepsilon+-\alpha T}{T}} \right)}=\\
    =\frac{e^{S_1} \left(\gamma_L \alpha_+ f(\varepsilon+\alpha T)+\gamma_R \alpha_- f(\varepsilon-\alpha T) \right)}{\gamma_L \alpha_+ f(\varepsilon+\alpha T)\left(e^{S_1}+e^{S_0}e^{\frac{\varepsilon}{T}} \right)+\gamma_R \alpha_- f(\varepsilon-\alpha T)\left(e^{S_1}+e^{S_0}e^{\frac{\varepsilon}{T}} \right)}=\\
    =\frac{e^{S_1}}{e^{S_1}+e^{S_0}e^{\frac{\varepsilon}{T}}}=\frac{1}{1+e^{\frac{\varepsilon-T\Delta S}{T}}}=f(\varepsilon-T\Delta S)
\end{multline}
It is an noteworthy and non-intuitive result that despite the induced bias voltage between the electrodes, the equilibrium population of the QD remains the same as for the standard case of obeyed detailed balance, and depends on the entropy difference between the charge states only. 

To find transport characteristics of the device, in line with Appendix \ref{app-GL}, we apply an infinitesimal change $\dd f$ to the equilibrium state of the left electrode. As a result, the stationary populations change to be equal to $P_1+\dd p$ and $P_0-\dd p$. This automatically satisfies the normalisation condition for a stationary state (Eq.\ref{eq-st-con}), while the second condition gives, to the linear order, as the equilibrium current is zero (we move to the asymmetry-based rates expressions -- Eq.\ref{eq-rates-DB}):
\begin{multline}
    \left((1+e^{\gamma})e^{S_1}(1+e^\alpha)P_0+(1+e^{\gamma})e^{S_0}(1+e^{-\alpha})P_1 \right)\dd f=\\
    =\left((1+e^{\gamma})e^{S_1}(1+e^\alpha)f_L+(1+e^{-\gamma})e^{S_1}(1+e^{-\alpha})f_R\right)\dd p +\\+
    \left((1+e^{\gamma})e^{S_0}(1+e^{-\alpha})(1-f_L)+(1+e^{-\gamma})e^{S_0}(1+e^{\alpha})(1-f_R) \right)\dd p
\end{multline}
Using $(1-f(\varepsilon))/f(\varepsilon)=e^{\varepsilon/T}$ and the properties of the choice of the rate coefficients, the above can be simplified to:
\begin{equation}
    \dd p=e^{-\Delta S}\frac{P_1^2}{f_L \left(e^\alpha f_L+e^{-\gamma} f_R \right)}\dd f
\end{equation}
Replacing $\dd f$ with the derivatives of the left Fermi distribution, we find the change of QD population with the application of incremental voltage and temperature differences:
\begin{equation}
    \begin{dcases}
        \frac{\dd p}{\dd V}=\frac{1}{T}e^{-\Delta S}\frac{P_1^2}{\left(e^\alpha f_L+e^{-\gamma} f_R \right)}(1-f_L)\\
        \frac{\dd p}{\dd T}=\frac{\varepsilon}{T^2}e^{-\Delta S}\frac{P_1^2}{\left(e^\alpha f_L+e^{-\gamma} f_R \right)}(1-f_L)
    \end{dcases}
\end{equation}
As before (Appendix \ref{app-GL}), the condutance and thermoelectric susceptibility can be written as:
\begin{equation}
    \begin{dcases}
        G=\frac{\dd I}{\dd V}=(\Gamma_{FR}+\Gamma_{TR})\frac{\dd p}{\dd V}\\
        L=\frac{\dd I}{\dd T}=(\Gamma_{FR}+\Gamma_{TR})\frac{\dd p}{\dd T}
    \end{dcases}
\end{equation}
Substituting the rates expressions (Eq.\ref{eq-rates-DB}) into $\Gamma_{FR}+\Gamma_{TR}$:
\begin{multline}
    \Gamma_{FR}+\Gamma_{TR}=\Gamma (1+e^{-\gamma})\left(e^{S_1}(1+e^{-\alpha})f_R+e^{S_0}(1+e^{\alpha})(1-f_R)\right)=\\
    = \Gamma e^{S_1} (1+e^{-\gamma}) \left((1+e^{-\alpha})f_R+ e^{-\Delta S}(1+e^{\alpha})e^{\frac{\varepsilon-\alpha T}{T}}f_R\right)=\\
    =\Gamma e^{S_1} (1+e^{-\gamma})(1+e^{-\alpha})\left(1+e^{\frac{\varepsilon-T\Delta S}{T}} \right)f_R=\Gamma e^{S_1} (1+e^{-\gamma})(1+e^{-\alpha})\frac{f_R}{P_1}
\end{multline}
Combining this with the expression for $\dd p/\dd V$, we find conductance, $G(\varepsilon)$ ($L=\varepsilon G/T$):
\begin{equation}
    G(\varepsilon)=\frac{\Gamma e^{S_0}}{T} P_1 \frac{(1+e^{-\gamma})(1+e^{\alpha})}{\left(e^\alpha f_L+e^{-\gamma} f_R \right)}e^{\frac{\varepsilon}{T}} f_L f_R
\end{equation}
This result is reminiscent of the conductance in Eq.\ref{eq-GL}. $P_1$ is equal to $f(\varepsilon-T\Delta S)$, while the weighted harmonic sum term $f_L f_R/(a f_L + f_R)$ is expected to be similar to a Fermi distribution, which, multiplied by $e^{\frac{\varepsilon}{T}}$, will give $1-f$ shifted by some offset energy. 

We now quantify these intuitive considerations by analysing the term:
\begin{equation}
\label{eq-SI-almost}
    \frac{(1+e^{-\gamma})(1+e^{\alpha})e^{\frac{\varepsilon}{T}} f_L f_R}{\left(e^\alpha f_L+e^{-\gamma} f_R \right)}=\frac{(1+e^{-\gamma})(1+e^{\alpha})e^{\frac{\varepsilon}{T}}}{e^\alpha (1+e^{\frac{\varepsilon-\alpha T}{T}})+e^{-\gamma} (1+e^{\frac{\varepsilon+\alpha T}{T}})}
\end{equation}
It is in fact it can be proved for a general case that a  ``weighted harmonic sum'' of two Fermi distributions is a Fermi distribution:
\begin{equation}
    \frac{f(\varepsilon-\mu)f(\varepsilon-\nu)}{Mf(\varepsilon-\mu)+Nf(\varepsilon-\nu)}=Lf(\varepsilon-\lambda)
\end{equation}
We, however, will demonstrate it for our specific case, while the general case proceeds similarly. The denominator in Eq.\ref{eq-SI-almost} can be modified as:
\begin{multline}
    e^\alpha (1+e^{\frac{\varepsilon-\alpha T}{T}})+e^{-\gamma} (1+e^{\frac{\varepsilon+\alpha T}{T}})=(e^\alpha+e^{-\gamma})\left(1+\frac{1+e^{\alpha-\gamma}}{e^\alpha+e^{-\gamma}}e^{\frac{\varepsilon}{T}} \right)=\\=
    (e^\alpha+e^{-\gamma})\left(1+e^{\frac{\varepsilon}{T}+\ln \left(\frac{1+e^{\alpha-\gamma}}{e^\alpha+e^{-\gamma}} \right)} \right)=(e^\alpha+e^{-\gamma})\frac{1}{f\left(\varepsilon+T\ln \left(\frac{1+e^{\alpha-\gamma}}{e^\alpha+e^{-\gamma}} \right) \right)}
\end{multline}
Substituting this into Eq.\ref{eq-SI-almost} gives:
\begin{equation}
    \frac{(1+e^{-\gamma})(1+e^{\alpha})e^{\frac{\varepsilon}{T}} f_L f_R}{\left(e^\alpha f_L+e^{-\gamma} f_R \right)}=\left(1+\frac{e^\alpha+e^{-\gamma}}{1+e^{\alpha-\gamma}} \right)\left(1-f\left(\varepsilon+T\ln \left(\frac{1+e^{\alpha-\gamma}}{e^\alpha+e^{-\gamma}} \right) \right) \right)
\end{equation}
Substituting the above into the expression for conductance, we arrive at the final result: 
\begin{equation}
    G(\varepsilon)=\frac{\Gamma}{T}e^{S_0} \left(1+\frac{e^{\alpha}+e^{-\gamma}}{1+e^{\alpha-\gamma}} \right) f(\varepsilon-T\Delta S)\left(1-f\left(\varepsilon-T \ln\frac{e^{\alpha}+e^{-\gamma}}{1+e^{\alpha-\gamma}}\right) \right) 
\end{equation}


\end{document}